\documentclass{article}
\usepackage{arxiv}

\usepackage{graphicx}
\usepackage{amsmath}
\usepackage{amssymb}
\usepackage{tabularx}
\usepackage{pbox}
\usepackage{hyperref}
\usepackage{caption}
\usepackage{geometry}
\usepackage{array}
\usepackage{booktabs}
\usepackage{setspace}
\usepackage{lineno}

\graphicspath{{./pix/}}
\newcommand{\includegraphicsdpi}[3]{
    {
    \pdfimageresolution=#1  
    \includegraphics[#2]{#3}
    }
}

\newcommand{\FigureSize}[4]{
    \begin{figure}
        \begin{center}
            \includegraphics[height=#2]{#1}
        \end{center}
        \caption{\label{fig:#3} #4}
    \end{figure} 
}

\newcolumntype{C}{>{\centering\arraybackslash}X} 

\title{A real-time UAS hyperspectral anomaly detection system}

\date{February 13, 2026}

\newif\ifuniqueAffiliation
\uniqueAffiliationtrue

\ifuniqueAffiliation 
\author{ \href{https://orcid.org/0000-0002-0379-8367}{\includegraphics[scale=0.06]{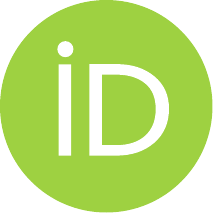}\hspace{1mm}Thomas P.~Watson} \\
	Electrical and Computer Engineering Dept. \\
	University of Memphis,\\
	Memphis, TN, USA \\
	\texttt{tpwatson@memphis.edu} \\
	\And
	Kevin McKenzie \\
	Electrical and Computer Engineering Dept. \\
	University of Memphis,\\
	Memphis, TN, USA \\
	\And
	Joseph Conroy \\
	Army Research Laboratory \\
	University of Memphis,\\
	Adelphi, MD, USA \\
	\And
	\href{https://orcid.org/0000-0002-6304-8381}{\includegraphics[scale=0.06]{orcid.pdf}\hspace{1mm}Eddie L. Jacobs} \\
	Electrical and Computer Engineering Dept. \\
	University of Memphis,\\
	Memphis, TN, USA \\
	\texttt{eljacobs@memphis.edu} \\
}
\else
\usepackage{authblk}

\newbox{\orcid}\sbox{\orcid}{\includegraphics[scale=0.06]{orcid.pdf}} 
\author[1]{%
	\href{https://orcid.org/0000-0000-0000-0000}{\usebox{\orcid}\hspace{1mm}David S.~Hippocampus\thanks{\texttt{hippo@cs.cranberry-lemon.edu}}}%
}
\author[1,2]{%
	\href{https://orcid.org/0000-0000-0000-0000}{\usebox{\orcid}\hspace{1mm}Elias D.~Striatum\thanks{\texttt{stariate@ee.mount-sheikh.edu}}}%
}
\affil[1]{Department of Computer Science, Cranberry-Lemon University, Pittsburgh, PA 15213}
\affil[2]{Department of Electrical Engineering, Mount-Sheikh University, Santa Narimana, Levand}
\fi

\begin{document}

\maketitle

\begin{abstract}
Detecting anomalies in hyperspectral image data, i.e. regions which are spectrally distinct from the image background, is a common task in hyperspectral imaging. Such regions may represent interesting objects to human operators, but obtaining results often requires post-processing of captured data, delaying insight. To address this limitation, we apply an anomaly detection algorithm to a visible and near-infrared (VNIR) push-broom hyperspectral image sensor in real time onboard a small uncrewed aerial system (UAS), exploring how UAS limitations affect the algorithm. As the generated anomaly information is much more concise than the raw hyperspectral data, it can feasibly be transmitted wirelessly. To detection, we couple an innovative and fast georectification algorithm that enables anomalous areas to be interactively investigated and characterized immediately by a human operator receiving the anomaly data at a ground station. Using these elements, we demonstrate a novel and complete end-to-end solution from data capture and preparation, through anomaly detection and transmission, to ground station display and interaction, all in real time and with relatively low cost components.

\end{abstract}

\section{Introduction}

\subsection{Background and Motivation}

Hyperspectral imaging~\cite{yang_ccd_2003} is a technique which records intensity values of hundreds of narrow portions (or bands) of a wide wavelength spectrum for each imaged pixel sample. This compares to multispectral imaging, which records weighted intensities of a few (e.g. three for the ubiquitous red-green-blue (RGB) camera) wide portions of a spectrum, and monochromatic imaging, which records only one weighted intensity value for a complete spectrum.

The wealth of information provided by a hyperspectral image is useful in a number of scenarios, including precision agriculture~\cite{ram_systematic_2024,lu_recent_2020}, environmental monitoring~\cite{stuart_hyperspectral_2019, arias_mapping_2025}, defense~\cite{shimoni_hypersectral_2019}, and other remote sensing applications. Deployment of a hyperspectral imager on an uncrewed aerial system (UAS) supports many of these applications~\cite{hasler_-it-yourself_2024}, but storing, processing, and leveraging the large volumes of data generated by the imager poses unique challenges~\cite{melian_real-time_2021}.

Hyperspectral anomaly detection~\cite{su_hyperspectral_2022} is a task that finds regions of the image that are spectrally distinct from a scene's background and thus may represent interesting objects. This detection requires no knowledge of the scene or potential objects and therefore is useful in many contexts without adjustment~\cite{chhapariya_target_2024}. However, large hyperspectral data volumes are impractical to transmit in real time from the UAS~\cite{melian_real-time_2021}. Traditional detection systems require data to be downloaded off the UAS after flight then post-processed on powerful machines~\cite{dastranj_remix_2025}, making immediate action on anomalies impossible.

Our work, to our knowledge, is the first to publicly demonstrate any hyperspectral anomaly detection algorithm operating in real-time onboard a UAS. We use an onboard computer to detect anomalies and a transmission system to send results on a continuous basis for analysis on the ground while the UAS is still in the air. This enables insight to begin immediately, before the UAS even completes take-off, opening up new applications for detection, observation, and response~\cite{scafutto_monitoring_2025,garske_erx_2025}. Our system can also support other algorithms, making it adaptable to precision agriculture and environmental monitoring.

\subsection{Related System Work}
Hyperspectral processing onboard a UAS has been studied before, and there is other work on systems that share steps in our processing chain (or perform related steps) with the overall goal of real-time processing. However, none that we know of combine and test all necessary components for real-time anomaly detection together in one system.

We summarize a variety of systems from the literature in Table~\ref{tab:related}, then discuss each system relative to ours below. The platform lists the vehicle type and processing device to give an idea of required size and available computational power. Values/sec estimates design data throughput in terms of hyperspectral bands per sample, times lines per second, times samples per line (BxLxS). Finally, the in-air steps describe which parts of the system's processing are performed by the aerial platform, as some systems split processing between air and ground.

\begin{table}
\caption{Onboard real-time hyperspectral processing system comparison \label{tab:related}}
\begin{tabularx}{\linewidth}{Ccccc}
\toprule
\textbf{Work} & \textbf{Platform} & \textbf{Values/sec} & \textbf{In-Air Steps} \\
\midrule

HyperLCA~\cite{melian_real-time_2021}, 2021 & UAS, Jetson & 160x200x1024 & Lossy Compression, Transmission \\

AMMIS~\cite{xue_novel_2023}, 2023 & Airplane, Desktop PC & 250x160x2048 & Feature Extraction \\

Precision~\cite{neri_real-time_2024}, 2024 & Ground, Raspberry Pi & 50x50x700 & Calibration, Detection, Transmission \\

HySpex~\cite{loke_next-gen_2025}, 2025 & UAS, Not published & Not published & Calibration, Detection, Transmission \\

REMIX~\cite{dastranj_remix_2025}, 2025 & UAS, Raspberry Pi & 3x100x682 & Calibration, Detection, Georectification \\

Concurrent~\cite{caba_concurrent_2025}, 2025 & UAS, FPGA & 160x200x1024 & Lossy Compression, Detection \\

Tandem~\cite{yang_onboard_2025}, 2025 & UAS, FPGA & 270x200x640 & Calibration \\

DPSR~\cite{piccinini_onboard_2025}, 2025 & Satellite, Jetson & 66x230x1000 & Image Enhancement \\

Ours & UAS, UP Squared Pro & 70x249x900 & Calibration, Detection, Transmission \\

\bottomrule
\end{tabularx}
\end{table}

Out of the systems in Table~\ref{tab:related}, the HyperLCA~\cite{melian_real-time_2021}, Tandem~\cite{yang_onboard_2025}, and DPSR~\cite{piccinini_onboard_2025} systems are designed for tasks other than anomaly detection. HyperLCA in particular recognizes the high bandwidth of hyperspectral imagery and the need to reduce it, like our system, but performs this via lossy compression instead of anomaly detection, and its transmission system is not designed for long range. The Concurrent~\cite{caba_concurrent_2025} system performs aerial anomaly detection, but doesn't discuss any mechanism to transmit the results. The AMMIS~\cite{xue_novel_2023} system performs feature extraction in the air, but actual anomaly detection on the ground, and requires powerful hardware on both sides. The remaining systems, Precision, HySpex, and REMIX, are much closer in structure and capability to our system, but they still lack our key abilities of full hyperspectral anomaly detection and immediate ground analysis.

The Precision~\cite{neri_real-time_2024} system is designed to distinguish between lettuce and arugula vegetable leaves from an uncrewed ground vehicle in a precision agriculture application. The system is lightweight enough to be adapted to UAS, but its detector is trained for only these two types of vegetables, and therefore is not an anomaly detector. There is also little published information on its transmission system, so we are unable to evaluate what results are available immediately on the ground.

The HySpex~\cite{loke_next-gen_2025} system has almost no published information, other than mentions of an optional transmission system and an implementation of anomaly detection. There is no data available on any detection algorithm, parameters, or capabilities. We are therefore unable to determine if the anomaly detection fully leverages the hyperspectral data, or confirm what results may be available immediately on the ground.

The REMIX~\cite{dastranj_remix_2025} system encompasses a framework for many tasks, though in that work it is applied to the task of harmful algae bloom detection. While the system uses a hyperspectral camera, the detector for this task leverages only three spectral bands and is designed specifically for algae blooms, so it is not an anomaly detector. The published performance figures suggest their system may be unable to scale to processing many more bands if necessary for a different task. The system also does not have a transmission component, so although most information is indeed processed in real-time onboard the UAS, landing and (short) post-processing is still required to access any results.

\subsection{Related Detection Work}

A classic algorithm for the anomaly detection task is the Reed-Xiaoli~\cite{reed_adaptive_1990} detection algorithm, which assigns an anomaly score to each pixel based on its probability within the statistical distribution of observed pixels, thereby judging small probabilities as anomalous. RX and its relatives, including nonlinear kernel RX~\cite{hidalgo_efficient_2021}, have shown good performance in a variety of applications. Therefore, we believe RX is a good candidate for our real-time system, and other studies have shown that more complex detectors may be too slow~\cite{garske_erx_2025,zhao_global_2015,li_hyperspectral_2024}.

Variants have been widely explored in the literature~\cite{zhao_global_2015}, including their application for real-time contexts. A well-known improvement of the RX algorithm is the kernel RX~\cite{kwon_kernel_2005} variant, which uses a nonlinear mapping kernel to improve the RX algorithm effectiveness, at the cost of performance for computation of the kernel.

Many other techniques to improve the performance of these algorithms have been demonstrated~\cite{hidalgo_efficient_2021, rossi_rx_2014}, including taking advantage of modern hardware parallelism~\cite{garzon_anomaly_2012}, simplifying the computations through progressive updates~\cite{garske_erx_2025}, recursive processing to take advantage of spatial information~\cite{he_recursive_2023}, and moving-window processing of the image to exploit data redundancy~\cite{zhao_fast_2018}.

\subsection{Our Contribution}

Our previous work~\cite{watson_evaluation_2023} tested several versions of the RX algorithm and showed that the simpler variants are sufficient for modern processors, yet still have good detection performance. We have also demonstrated~\cite{watson_spacecube_2026} real-time approaches to the calibration and georectification steps.

This work builds on those by describing and testing a complete system using the algorithms and lessons we previously learned. We perform real-time anomaly detection, including calibration~\cite{k_c_lawrence_calibration_2003} to compensate for sensor non-idealities, georectification~\cite{warren_data_2014} to compensate for UAS motion, and develop a transmission system and ground user interface to leverage the results. We show that the classic RX algorithm effectively summarizes the hyperspectral data, demonstrate that the detection is useful visually and matches well compared to a ground truth, and confirm the required processing is feasible in real-time.

In Section~\ref{sec:architecture}, we describe our system architecture, including a UAS, hyperspectral sensor, onboard computer, detection algorithm, transmission system, and ground control station (GCS). In Section~\ref{sec:implementation}, we detail our system's implementation, which can capture hyperspectral data from the sensor, perform anomaly detection on the air data, then transmit the results to the ground for interactive investigation. In Section~\ref{sec:methodology}, we collect data of staged anomalies using our system, and in Section~\ref{sec:results} we present those results to evaluate performance and behavior. Finally, in Section~\ref{sec:discussion} we discuss the results and evaluation, including some system limitations and possible improvements.

\section{System Architecture} \label{sec:architecture}

\begin{figure}
\centering
\includegraphics[width=6.5in]{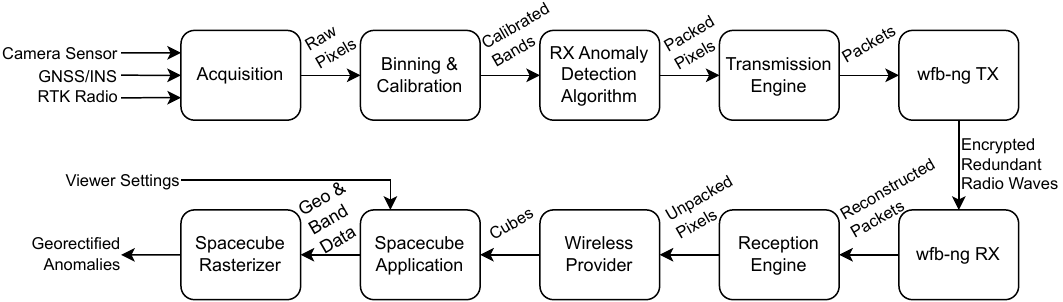}
\caption 
{ \label{fig:system_diagram}
System data flow } 
\end{figure}

Our system starts with light entering the hyperspectral sensor on a UAS and ends with anomaly information being displayed to the operator on the ground. In this section we explain the system's background and organization, including the steps within, with a particular focus on algorithm selection and the division of processing. The overall data flow of the system is illustrated in Figure~\ref{fig:system_diagram}. In Section~\ref{sec:implementation}, we discuss the specific implementation of the steps.

\subsection{Data Acquisition} \label{subsec:arch_acq}

Our system is designed for line-scan hyperspectral imagers~\cite{yang_ccd_2003}, which capture one line of pixels at a time. For each pixel, a spectrum is captured that quantifies the intensity of the light at each particular wavelength band. The system is mounted on a UAS so that the line captured is perpendicular to the direction of flight. The UAS's flight then changes the position of the line through time, allowing a complete image to be built with spectral information for each pixel.

As the UAS does not move in a precisely straight line, we also need to know where the imager is at each line, using the data to assemble a coherent image in a process known as georectification~\cite{warren_data_2014,angel_automated_2019}. A combined Global Navigation Satellite System (GNSS) and Inertial Navigation System (INS) tracks, computes, and reports the position and orientation of the imager to the acquisition system.

The imager's sensor is not equally sensitive at each pixel and band, and the sensor settings (such as exposure time and gain) are continually adjusted to optimally capture each line. To eliminate distortion from these effects, the pixel data is scaled according to the sensor settings, then a calibration is applied to convert these values into radiometric quantities~\cite{k_c_lawrence_calibration_2003}. This so-called radiance data is independent of the imager, sensor, and configuration, making it suitable for processing.

Many applications of hyperspectral imagery further remove the effect of light on the scene, thereby isolating purely the materials in it, by computing reflectance~\cite{geladi_hyperspectral_2004} from the radiance data. However, this requires knowing the light incident upon the scene by using a target of known reflectance, another sensor to measure it directly, or known weather and a corresponding model. Fortunately, as the reflectance correction is a linear operation performed identically on all pixels, it does not change the results of the linear RX detector (selected in Section~\ref{subsec:architecture_algo}). The step can therefore be safely omitted, reducing complexity.

Though too large to transmit to the ground, the full raw data captured from the sensor, plus associated settings and GNSS/INS track, is saved to disk onboard the aerial system for post-processing and analysis. The transmitted data packets and timestamps are also saved for debugging and replay.

\subsection{Detection Algorithm} \label{subsec:architecture_algo}

From the variety of RX variants, we must pick a specific implementation. Our prior work~\cite{watson_evaluation_2023} tested data from our system's hyperspectral imager on different variants with different configurations to understand the limitations of our system's computing platform. The relevant results are repeated here and used to inform our algorithm choice.

\begin{figure}
    \begin{center}
        \includegraphicsdpi{300}{width=10cm}{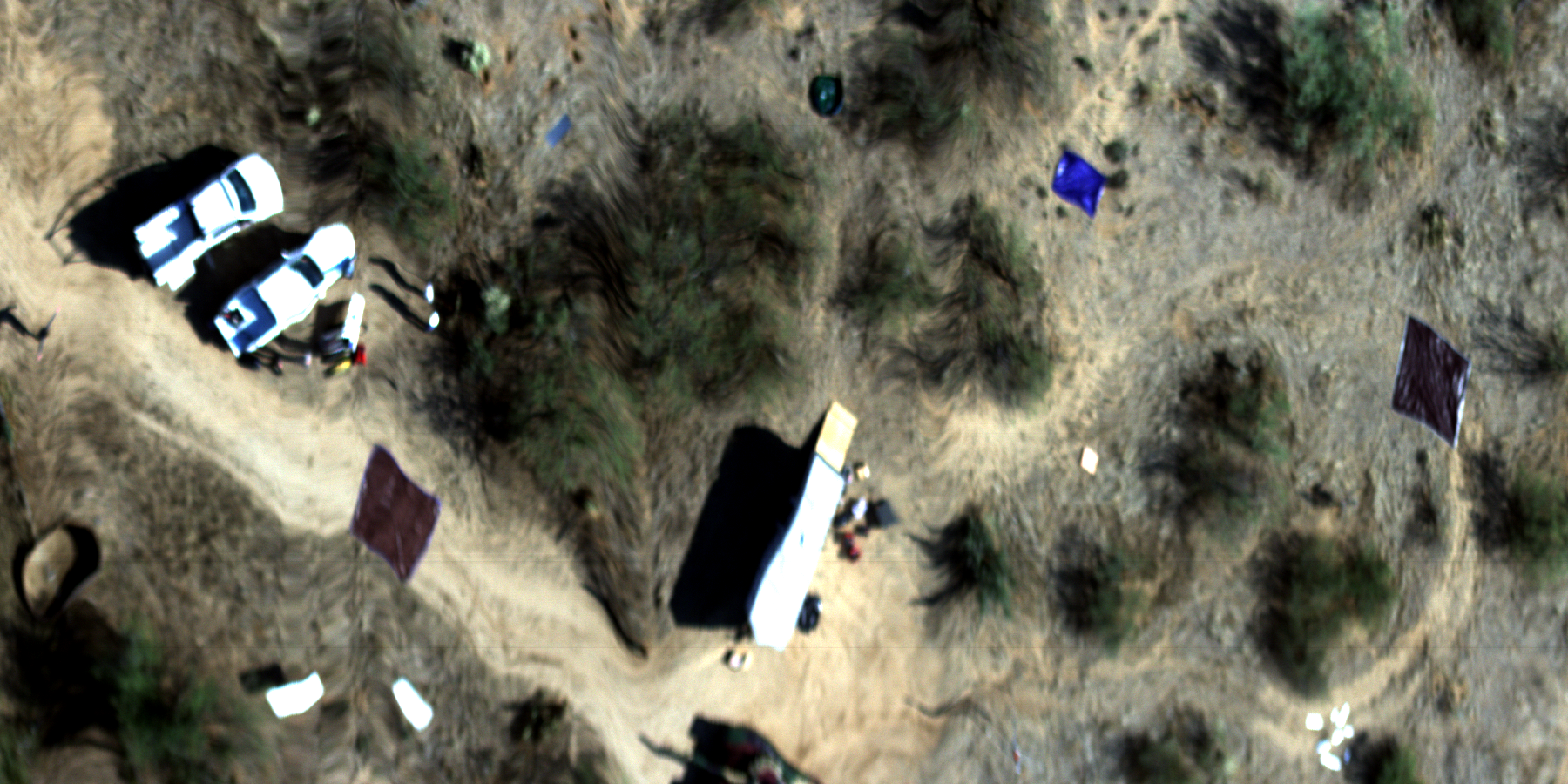}
    \end{center}
    \caption{\label{fig:orig_cube} Anomaly test scene}
\end{figure}

In that work, we staged a scene at the Santa Rita Experimental Range owned by the University of Arizona in December 2022. The tested data is shown in Figure~\ref{fig:orig_cube}; the lack of georectification is visible as wiggles as the flight proceeds from left to right. The data was acquired over a 7.2 second period and consists of 1800 vertical lines from left to right, 900 samples in each line from top to bottom, and 300 total bands. For display, the bands closest to red, green, and blue are extracted.

\begin{table}
\caption{Anomaly detection algorithm results \label{tab:results}}      
\begin{tabularx}{\textwidth}{cccc}
\toprule
\textbf{Algorithm} & \textbf{Dimensions} & \textbf{Processing Time} & \textbf{AUC} \\
\midrule

Global RX & 1800 $\times$ 900 $\times$ 300 & 32.24s ± 0.36s & 0.62 \\

3x Binned Global RX & 1800 $\times$ 900 $\times$ 100 & 7.00s ± 0.06s & 0.71 \\

Subsampled Kernel RX & 1800 $\times$ 900 $\times$ 300 & 5.77s ± 0.05s & 0.78 \\

Georectified Global RX & 849 $\times$ 291 $\times$ 300 & 4.24s ± 0.01s & 0.60 \\

Georectified Subsampled Kernel RX & 849 $\times$ 291 $\times$ 300 & 0.87s ± 0.01s & 0.83 \\

\bottomrule
\end{tabularx}
\end{table}

The results from relevant tested RX variations are presented in Table~\ref{tab:results} as total processing time (average ± standard deviation of 10 runs) and accuracy (Receiver Operating Characteristic Area under Curve (AUC)~\cite{fawcett_introduction_2006}). The full receiver operating characteristic curves are presented in Figure~\ref{fig:results_final}. An ideal detector would have a curve which touched the top and left borders of the graph and an AUC of 1. A random detector would show a diagonal curve and an AUC of 0.5.

\FigureSize{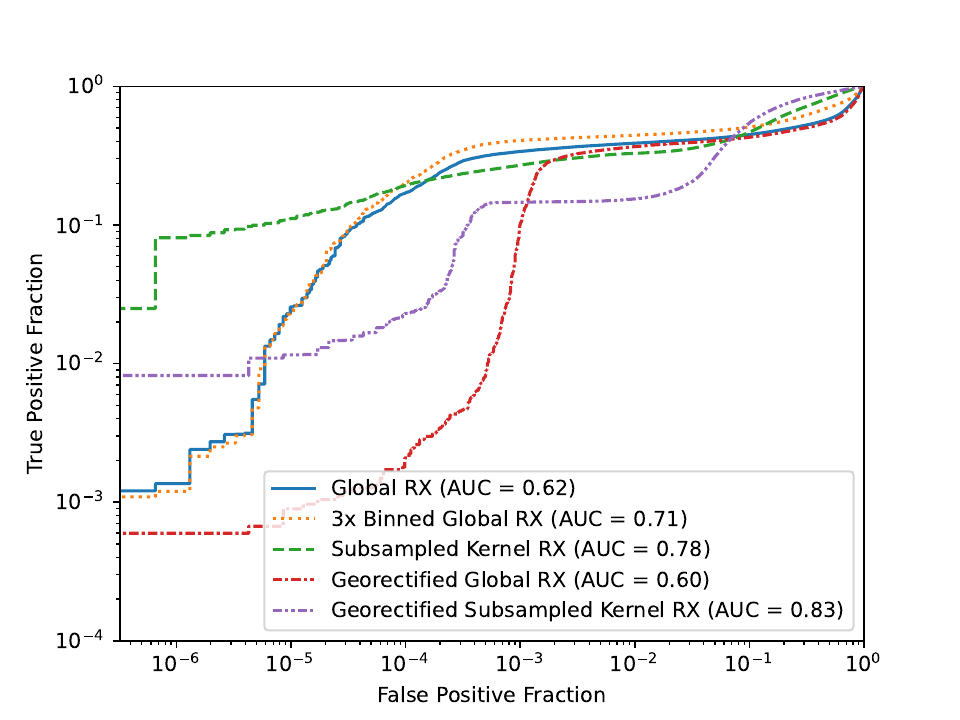}{8cm}{results_final}{Detector Receiver Operating Characteristic (ROC) curves}

The data show that the classic global RX~\cite{reed_adaptive_1990} algorithm (where the distribution is computed over all pixels in the image) has moderate accuracy, but is too slow for real-time operation unless binned to 3x or more (i.e. 3 bands are summed into 1). The binning also increases accuracy, likely by reducing noise from the sensor. A more advanced variant, subsampled kernel RX~\cite{hidalgo_efficient_2021}, is faster and nominally higher accuracy, but curiously shows a lower true positive rate at reasonable false positive rates (between $10^{-4}$ and $10^{-1}$), calling into question the actual benefit.

Performing georectification before detection makes intuitive sense and simplifies implementation by significantly reducing the amount of data the detector processes and the radio transmits. Unfortunately, this poses a severe limitation as georectification settings can no longer be changed by the operator because the ground station does not have the source data.

The RX algorithm does not consider spatial relationships between pixels, so it is not in principle affected by the presence of georectification or its accuracy. It is of course possible for anomalous objects to be artificially enlarged or shrunk by platform motion and so change their portion of the overall distribution, but evidently this has only a small effect on the accuracy.

Based on our previous results and in light of the full system implementation, we chose the simple and classic binned global RX algorithm, and performed it before georectification. Though we focus on this variant of RX, our system's architecture is largely independent of the algorithm and its task. Any algorithm which can produce a score for each acquired pixel within the available processing constraints could be integrated, including more sophisticated anomaly detectors. Another task could include calculation of a selected agricultural plant health metric~\cite{lu_recent_2020}, allowing an operator to immediately see the most important information about a crop field.

\subsection{Data Transmission}

Despite the massive reduction in data volume made possible by computing anomaly scores in the air and transmitting one value per pixel instead of 300, the required transmit bandwidth is still several megabits per second, much higher than the kilobits per second available on standard telemetry radios~\cite{ardupilot_project_sik_2025}. We need something with more bandwidth, yet would also like a long communication range so the UAS can survey a large area.

WiFi can offer very high bandwidths, but it usually does not have sufficient range due to protocol limitations~\cite{asadpour_ground_2013}. Cellular radios would offer good bandwidth and essentially infinite effective range, but they require connection to a cell tower which might not be available in remote regions, and a subscription to a data plan which represents an ongoing cost. Internet Protocol (IP) radios (e.g. Doodle Labs~\cite{doodle_labs_mini-oem_2024}) combine strengths of WiFi and cellular, offering relatively high bandwidth at relatively long ranges without other infrastructure, but they are expensive special purpose items.

These higher bandwidth systems also generally layer IP communications (IP addresses, sockets, TCP, etc.) on top of the radio link, including automatic routing and link adaptation, so that the bandwidth, latency, and data path varies according to channel conditions. They unsurprisingly operate well in IP-network-based systems, but don't offer control; their adaptation schemes can reduce bandwidth or increase latency arbitrarily to the detriment of system function~\cite{fabra_methodology_2017}. Serial telemetry radios operate point-to-point, such that data transmitted by one side is received by any other in range without routing, but portions may end up corrupted or missing as conditions change.

However, certain models of WiFi radios can be placed into so-called monitor mode~\cite{gunther_analysis_2014}, where packets are transmitted with configurable bandwidth, power, and encoding scheme, and where all (non-corrupt) packets heard are received. Radios in this mode bypass the range and link limits of the WiFi protocol and operate much like standard telemetry radios. This facilitates a very inexpensive, high bandwidth, and customizable point-to-point radio system~\cite{vanhoef_testing_2023}. Adaptation and routing can be ignored or implemented at the application level where necessary.

Due to the low cost and customization possibilities, we use monitor-mode WiFi to transmit all data from the aerial system to our GCS; there is no link from our GCS back to the aerial system. Our transmission scheme accepts the possibility of missing data at the receive side rather than reducing bandwidth and dropping data at the transmit side. This guarantees full system performance where conditions permit, simplifying design and providing a better user experience, though at the cost of failing to receive in challenging conditions instead of offering reduced performance.

\subsection{Ground Control Station} \label{subsec:gcs}

Any Linux system is suitable for the ground control station, though a customized driver is required for the WiFi hardware. The software is based on our previous work~\cite{watson_spacecube_2026}, Spacecube, which performs real-time hyperspectral georectification. One of Spacecube's major components is a provider library, so-named as it provides the data Spacecube processes, ordinarily from disk.

For our real-time system, we instead create another provider implementation that sources data from the WiFi radio. Spacecube's normal mechanisms are then used by the operator to pan and zoom the data, change viewing settings, and so forth. Spacecube georectifies the scene according to the current settings, immediately re-processing all collected data when the settings are changed. Re-processing would be impossible if the georectification were performed in the air.

Using Spacecube's interface, the operator can elect to display per-pixel anomaly scores (colored according to value), a binary image with a selectable threshold that displays if pixels are anomalous or not, or the scene in RGB colors for context.

\section{System Implementation} \label{sec:implementation}

After understanding the architecture described in Section~\ref{sec:architecture}, including capture, anomaly detection, transmission, and ground control/interaction, we now discuss specific implementation details of the components, including selection of hardware, choice of parameters, and description of algorithms. More information on specific components and operator interaction is available with the source code~\cite{watson_spacecube_2025}.

\subsection{Flight System}

For our system to operate aerially, we need some sort of platform to fly it on and device to perform the required computations. A lot of factors can go into selection of a UAS platform, but for our primary experiments we utilized a 1000mm-class hexcopter. The system was also partially tested on a fixed-wing UAS. Many factors can go into selection of an onboard computer platform as well. Ultimately, we decided on the UP Squared Pro~\cite{up_up_2023} with a 4 core Intel Pentium N4200 1.1GHz processor and 8GB of RAM. This SBC isn't very capable by desktop computer standards, but it has more CPU power and higher compatibility than alternatives like Raspberry Pi~\cite{raspberry_pi_raspberry_2023} or NVIDIA Jetson~\cite{nvidia_embedded_2023}.

The SBC's four cores are each used to perform one part of the processing chain. Performing other processes in parallel with continuous data acquisition defines a real-time system. The 8GB RAM allows buffering enough data to facilitate this parallelization. The standard USB interfaces allow attachment of our hardware. The standard Intel processor means that software and drivers compatible with ordinary PCs are also compatible with this SBC, unlike ARM-based systems such as Raspberry Pi. Compatibility speeds up experimentation and implementation to allow development of a more sophisticated system.

\begin{figure}
\centering
\includegraphics[height=6.0cm]{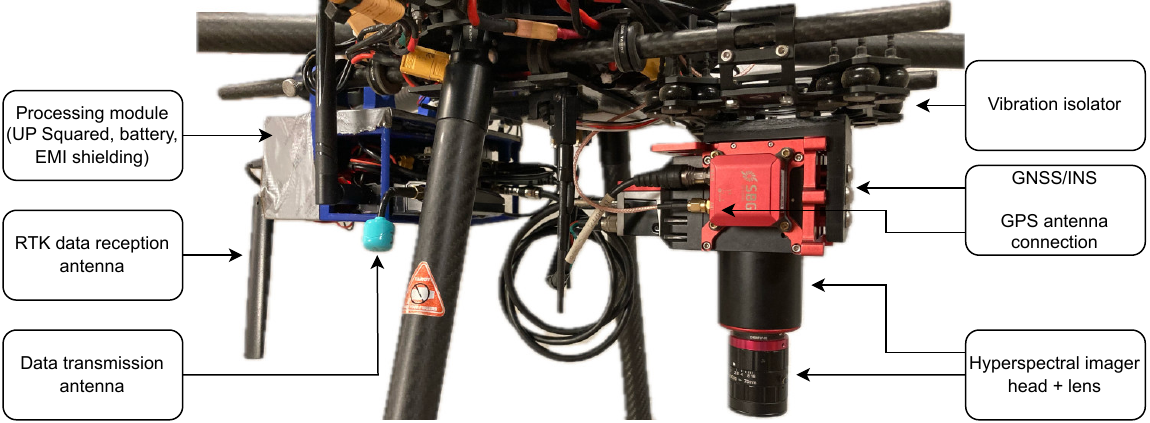}
\caption{\label{fig:drone_diagram} UAS and flight system (front on the right, two legs in the middle)}
\end{figure} 

The flight system is shown, attached to our UAS, in Figure~\ref{fig:drone_diagram}. The processing module is attached on the rear of the UAS (left in the image). The module is a 3D-printed assembly containing the SBC, a solid-state disk storing recordings, the radios, and a battery for system power. The module's exterior is mostly covered in electromagnetic interference shielding, secured with duct tape, to reduce dangerous interference with both the UAS's and the imager's GNSS. The hyperspectral imager (detailed in Section~\ref{subsec:acquisition}) is connected to USB ports on the front-facing side of the module. The system weight (including battery, imager, and mounting) is under 2.1kg, and the peak power consumption (with all components) is under 25W.

\subsection{Data Acquisition} \label{subsec:acquisition}

The hyperspectral imager we selected was the Resonon Pika L~\cite{resonon_resonon_2023}, primarily due to its light weight and low cost. This visible and near-infrared (VNIR) imager provides 300 bands which cover the 400-1000nm wavelength range (though approx. 20 are outside it), a nominal full-width half-maximum (FWHM) spectral resolution of 2.7nm, and (in our tests) a 47.5 degree field of view objective lens. To the imager is attached an SBG Systems Ellipse~N GNSS/INS that provides its position and orientation.

The imager is mounted to the front of the UAS (right side of Figure~\ref{fig:drone_diagram}) with the imaged line oriented nadir and perpendicular to the flight direction. It is bolted with a 3D-printed adapter to a vibration isolator to reduce the effect of mechanical noise from UAS propellers etc. on the capture. A helical GNSS antenna (not shown) is secured to the top of the UAS and connected via a cable to the INS.

The system captures 249 lines per second (249Hz). Each line is exposed for a fixed 3.9 milliseconds and contains 900 spatial samples and 300 wavelength bands. Gain is continually adjusted for each line to avoid saturation. The INS produces position and orientation data at 200Hz. The time of each line's start is measured with microsecond precision using a trigger pulse, allowing accurate interpolation.

The capture is performed in groups of 1000 lines, termed cubes. After each cube is collected (taking 4.02 seconds), the raw data is vidpak~\cite{watson_vidpak_2025}-compressed then written to disk. Out-of-specification bands below 400 nm and above 1000 nm are then discarded, leaving 280 bands; the discarded bands are also the noisiest and therefore least useful.

To meet processing time requirements of four seconds per cube, the data is binned by a factor of four, i.e. four consecutive bands in the original 280-band cube are summed to produce one band in the processed 70-band cube. This correspondingly reduces spectral resolution to roughly 9.1nm. It is possible that a factor of three would be sufficient reduction to process in real-time based on earlier results, but four gives a reasonable margin of safety.

In preparation for detection, the data is corrected for varying gain, then the radiometric calibration is applied using Spacecube's method. This converts the sensor's raw values, whose scale varies based on gain and position on the sensor, to radiance values, which represent a uniform response to illumination from the scene. This allows the detection algorithm to accurately analyze the scene without influence from the sensor settings or variance.

For operator context, an RGB color is determined for each pixel by retrieving the calibrated value from the band closest to each RGB wavelength of 640nm, 550nm, and 460nm respectively. While this closest-band approach does not produce the same result as the weighted wide-spectrum response of a typical RGB camera, it is simple, fast, and adequate for this auxiliary purpose.

\subsection{Detection Algorithm}

As discussed in Section~\ref{subsec:architecture_algo}, we use a classic ``global" RX implementation where each 1000-line cube is treated as an independent entity. In this algorithm, a multi-variate Gaussian distribution is computed over all pixels in the cube, then each pixel's anomaly score is computed as the squared Mahalanobis distance of that pixel's value from the distribution.

To compute the distribution, we treat each band as a variable and each pixel as a sample. Computing it using ``all pixels", of course, includes potential anomalies, and the significance of each pixel is independent of its location in the overall cube. The mean is computed as the average of each band independently, and the covariance matrix is calculated using standard statistical methods. For $B$ bands, the distribution is described by a $B$-length mean vector $\mu$ and a $B \times B$ covariance matrix $\Sigma$.

The score $\delta$ for each $B$-length vector pixel sample $s$ is then computed as

\begin{linenomath}
\begin{equation}
\delta = (s - \mu)^{T}\Sigma^{-1}(s - \mu)
\end{equation}
\end{linenomath}

The score increases as the probability of the given pixel within the distribution decreases. It is zero for a pixel equal to the mean, and increases inversely proportional to the variance in each dimension. Therefore, in bands with high variance where many pixels may be within the distribution, a large difference from the mean is required to get a high score. Similarly, with a low variance, a small difference is sufficient for a high score. The score is calculated across all bands to highlight pixels that are far away from the overall mean (i.e. improbable), and thus may be counted as anomalous in the scene.

This algorithm assumes that anomalous objects are infrequent to avoid skewing the distribution to the point that they are included. The algorithm can also account for larger anomalies that are very distinct. However, there is no method of tuning it for a particular anomaly, even if the anomaly's characteristics are known. To allow comparison across cubes, the scores are normalized so a value of 1 is the highest score in the cube. A threshold operation is also provided to mark as anomalous any pixels with a score above a configurable value.

The score is computed for each pixel then paired with its corresponding RGB color in preparation for transmission. The system devotes one processor core to the anomaly detection algorithm; others are used for other processing stages. Provided it can operate in real time with the available processing power, this score could easily be computed with a wide variety of algorithms, including more sophisticated anomaly detectors, or other functions for agricultural tasks.

\subsection{Data Transmission}

As mentioned in Section~\ref{subsec:gcs}, the information ultimately displayed to the operator is either the anomaly score or an RGB image of the scene. As the operator has the option to switch between them at any time, both are transmitted.

The system transmits every captured pixel, meaning 249 lines per second of 900 pixels each. Data compression is not used to ensure consistent performance and resilience to reception dropouts. Multiplying these values by the number of bits transmitted per pixel therefore results in the data rate of transmission. As operating at a higher data rate generally requires higher radio sensitivity, translating to a reduction in range for a given radio system, we should be parsimonious in our allocation of bits. 

We reduce the anomaly data size by transmitting using the IEEE-754 half precision floating point format so that each score is 16 bits, instead of the 32-bit single precision format used during calculation. As the data is intended for display to a human operator on a standard 8-bit monitor, a 16 bit value still has sufficient precision, even allowing for scaling and thresholding during display. As the squared Mahalanobis distance used for RX scoring is typically extremely skewed towards low values, we take the square root before transmission to make better use of the limited range. We also reduce the color data size using RGB565 format, where each color is allocated the corresponding number of bits. To optimize use of the limited bits, each color and score value is normalized to the cube's maximum before transmission.

This sums to 32 bits (or 4 bytes) per pixel, resulting in an image data rate of 7.17 megabits per second (Mbit/s). For comparison, the raw 12-bit band data would require 450 bytes per pixel for an image data rate of 807Mbit/s, which is non-trivial to transmit long distances.

Our radio, described in detail below, can transmit up to 4000 bytes of user data per packet, which is a convenient match to one line requiring $900 * 4 = 3600$ bytes of data. Each packet thus contains one line of data, plus a 128-byte header, resulting in transmission of 249 packets per second of 3728 bytes each, or a total data bandwidth of 7.43 Mbit/s.

The header contains cube ID and sequence information, allowing the cube to be set up and each line to be inserted into its proper place; the ground station fills missing lines in with black. It also contains the maximum value of each color and score so normalization can be undone on reception. For georectification, the header contains the two INS positions/orientations sampled before and after the start of exposure, plus exposure start time, so the GCS can interpolate the exact position of the start of each line. Many data fields are the same for all lines in the cube; this allows reconstruction of a valid (if mostly empty) cube if at least one line is successfully received.

To this data, we add 50\% forward error correction (FEC) information. After every 50 data packets transmitted by the application, an additional 25 packets containing parity information (generated using Reed-Solomon~\cite{wicker_reed-solomon_1999} coding) are also transmitted. If any 50 out of the 75 packets in the group are received, all 50 data packets can be reconstructed; otherwise only the actual data packets received can be used. This eliminates data loss in the presence of short reception gaps.

For actual data transmission, we rely on a slightly modified version of the wfb-ng~\cite{evseenko_wfb-ng_2023} program. It receives our packets to transmit over a Unix domain socket (instead of lossy UDP as in the original wfb-ng), generates the FEC data according to our 50+25 settings, adds encryption and authentication to ensure tamper resistance, then sends the packet data to a monitor-mode-enabled ALFA Network AWUS036ACH WiFi radio, which then pumps it out over the airwaves. One transmitted cube of 1000 lines is exactly 20 FEC groups, and including FEC and other overhead is transmitted at an average bandwidth of 11.4Mbit/s. 

The radio is configured for 20MHz bandwidth, 0.8$\mu$s guard interval, and Modulation Coding Scheme (MCS) index of 2, resulting in a peak transmission rate of precisely 19.5Mbit/s. Radio headers are transmitted at a lower rate, and transmission is not continuous due to contention and inter-packet gaps, so the actual sustainable rate is 10-20\% lower. This configuration provides good margin above our 11.4Mbit/s requirement, unlike the next slower rate. The radio broadcasts at a center frequency of 5.745GHz (WiFi channel index 149) and transmit power of 20 dBm (index 53) over an omnidirectional 5.8GHz antenna.

\subsection{Ground Control Station}

The ground control station (Figure~\ref{fig:gcs_deck}) is built around a Valve Steam Deck handheld gaming PC that incorporates a moderately powerful AMD Zen 2 2.4GHz CPU+GPU, 16GB of RAM, battery, and user interface (touchscreen, mouse touchpads, joysticks, buttons). The Steam Deck also has a USB port into which the receiving radio (ALFA Network AWUS036ACH, same as the transmitter) is plugged. The Steam Deck is an ergonomic system and provides easy mounting for the radio as well. The radio has one omnidirectional antenna and one linear antenna for diversity; both antennas can receive data.

\FigureSize{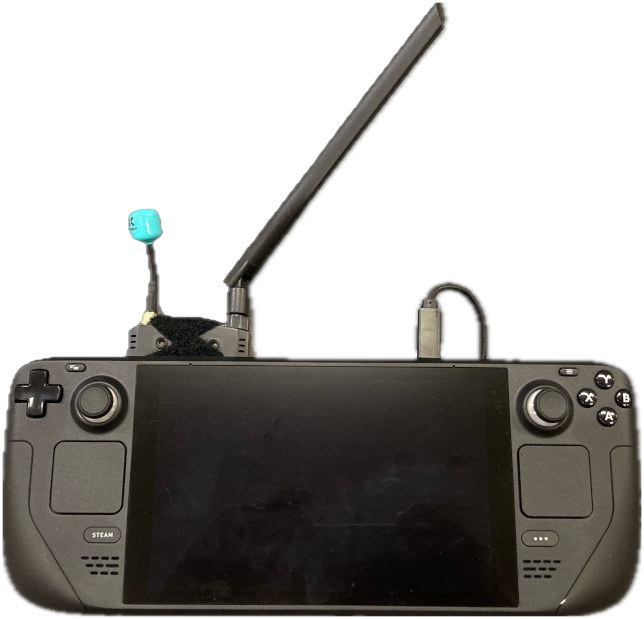}{4cm}{gcs_deck}{Ground control station with radio and antennas mounted}

The wfb-ng receiver runs on the GCS and listens to the radio for packets. It rejects packets which fail authentication due to tampering or corruption, decrypts successfully validated packets, and uses the FEC data to reconstruct missing packets as necessary. It then sends the packets over another Unix domain socket to the Spacecube provider.

The Spacecube provider organizes the received packets into cubes, unpacks the 32 bits of data per pixel (including undoing transmission normalization and squaring the scores), and uses the position and orientation information included in the packet headers to interpolate the start of each received line. The completed cubes are then provided to the rest of Spacecube for display. The provider also logs the received packets and timestamps for replay and testing.

\begin{figure}

    \begin{center}
        \includegraphics[width=16cm]{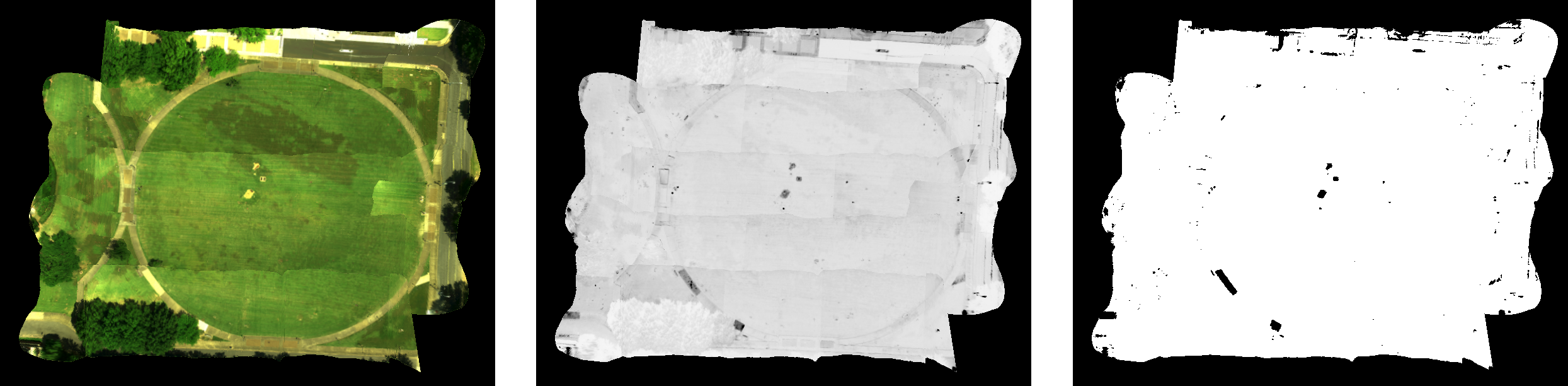}
    \end{center}
    \caption{\label{fig:ui} User interface modes (shown on fast flight without tank): calibrated RGB color (left), anomaly score (stretched 3x) with darker as more anomalous (center), anomaly threshold (0.110) with black as anomalous (right)}
\end{figure}

Spacecube performs real-time georectification of the data and allows the user to choose between RGB color, anomaly score, and thresholded anomalies (Figure~\ref{fig:ui}). The GCS touchscreen is used to pan and zoom the data. While the user interface is currently optimized for the anomaly detection task, appropriate options could easily be added for agriculture scores and tasks, including color maps and overlays.

\section{Experiment Methodology} \label{sec:methodology}

With the described and implemented system, we performed several flight tests to characterize its performance and behavior in various dimensions. In this section, we describe the flights and summarize the data we collected. In Section~\ref{sec:results} we analyze the collected data.

\subsection{Memphis Flights}
\begin{figure}
\centering
\includegraphics[width=3.49in]{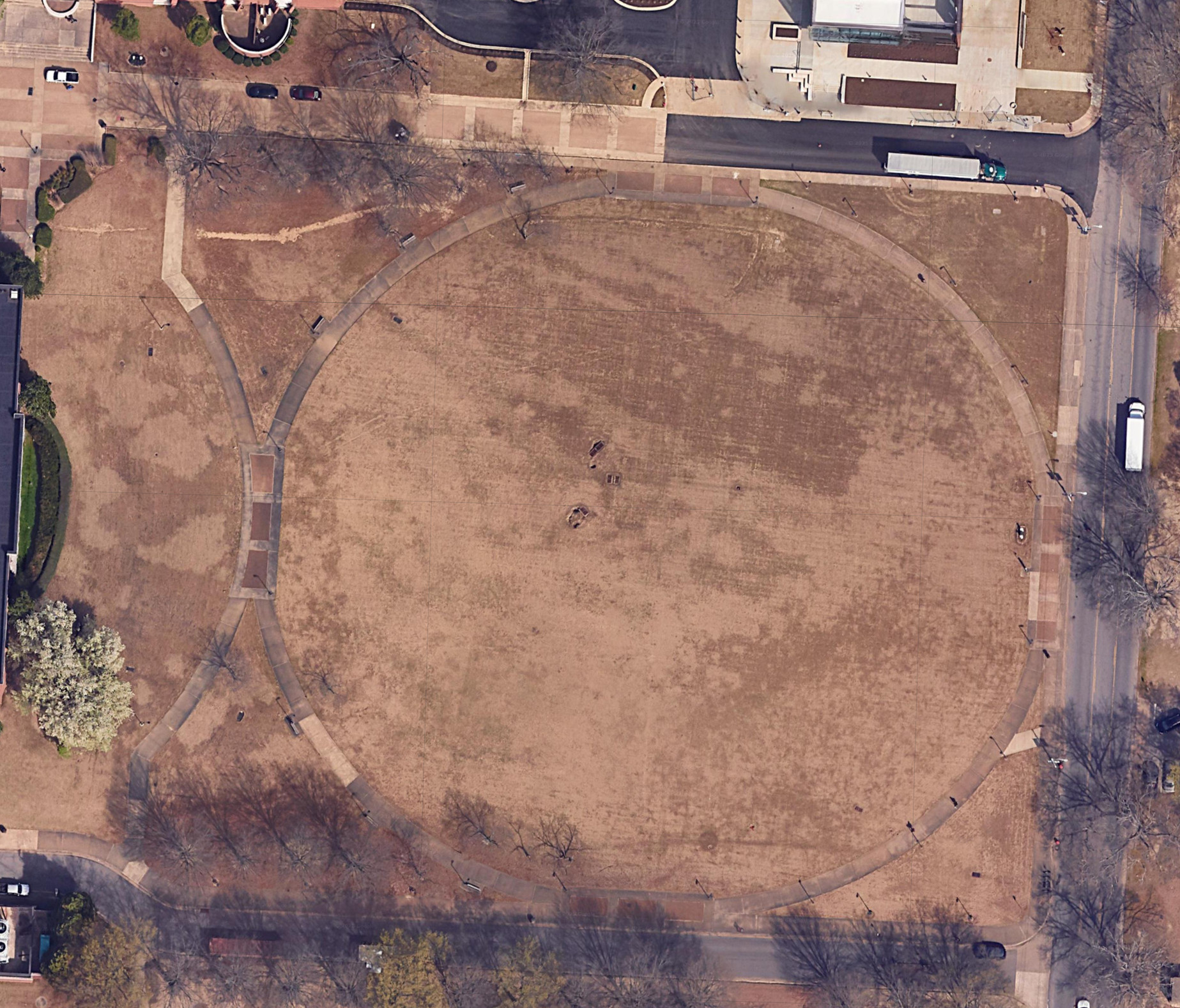}
\caption 
{ \label{fig:google_maps_image}
Captured Memphis area satellite imagery (Imagery/Map Data \copyright 2025 Google) } 
\end{figure}

In July 2025, we flew our primary test flights over a grassy area known as The Ellipse at the University of Memphis in Memphis, TN. We set up an inflatable tank to test detection of a relevant-to-us anomalous object. A representative satellite image of the area is shown in Figure~\ref{fig:google_maps_image}. 

We flew at a nominal height of 40m over two scenes (both the same area) and at two speeds (slow then fast), totaling four separate flights. The first scene did not include the tank and was recorded around 11:45 AM, while the second scene included the tank (and some supporting equipment) and was recorded around 3:25 PM. The slow speed was 5m/s and 7 horizontal passes with a capture duration (not including calibration and takeoff/return) of 4 minutes 13 seconds, while the fast speed was 10m/s and 5 horizontal passes with a capture duration of 2 minutes 12 seconds.

\begin{figure}
\centering
\includegraphics[width=3.49in]{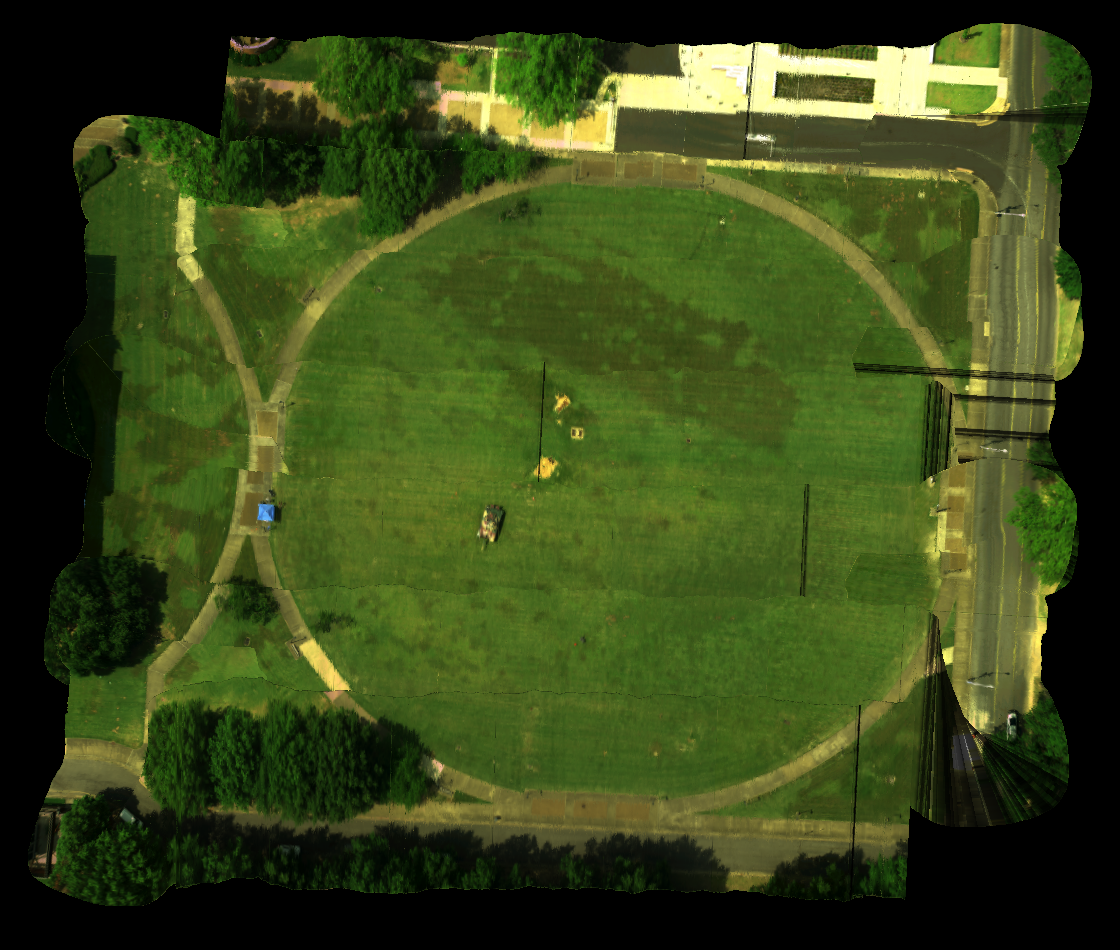}
\caption 
{ \label{fig:rgb_image}
Captured Memphis RGB image (slow flight with tank) } 
\end{figure}

Figure~\ref{fig:rgb_image} shows the RGB data captured by our system during the slow flight and including the tank. Black stripes on the right show missing lines that were lost in transmission. Operations were conducted from underneath the blue canopy in the left middle of the image (this canopy was not present in the non-tank scene). One operator held the Steam Deck GCS with the radio (including antennas) attached to its rear, while another operator held the UAS flight controller.

We tested that the interaction features worked properly during the flights, recorded the received data and processing times for analysis, and recorded the packet loss rate (after error correction) to characterize range. We also confirmed visually that the system latency was not too excessive (though 15-20 seconds was expected) and that it continued to perform well over the total flight duration.

\subsection{Uvalde Flight}

In May 2025, during the development of the system, we also flew over a staged scene in a desert environment in Uvalde, TX. The scene included two tanks (of different materials), plus other supporting vehicles and infrastructure for the operation. Unlike the Memphis flights, this flight was done with a fixed-wing UAS. The nominal height was 90m and nominal speed was 22m/s. The flight was performed around 11:30 AM, and included three horizontal passes and a capture duration of 5 minutes 42 seconds.

\begin{figure}
\centering
\includegraphics[width=5in]{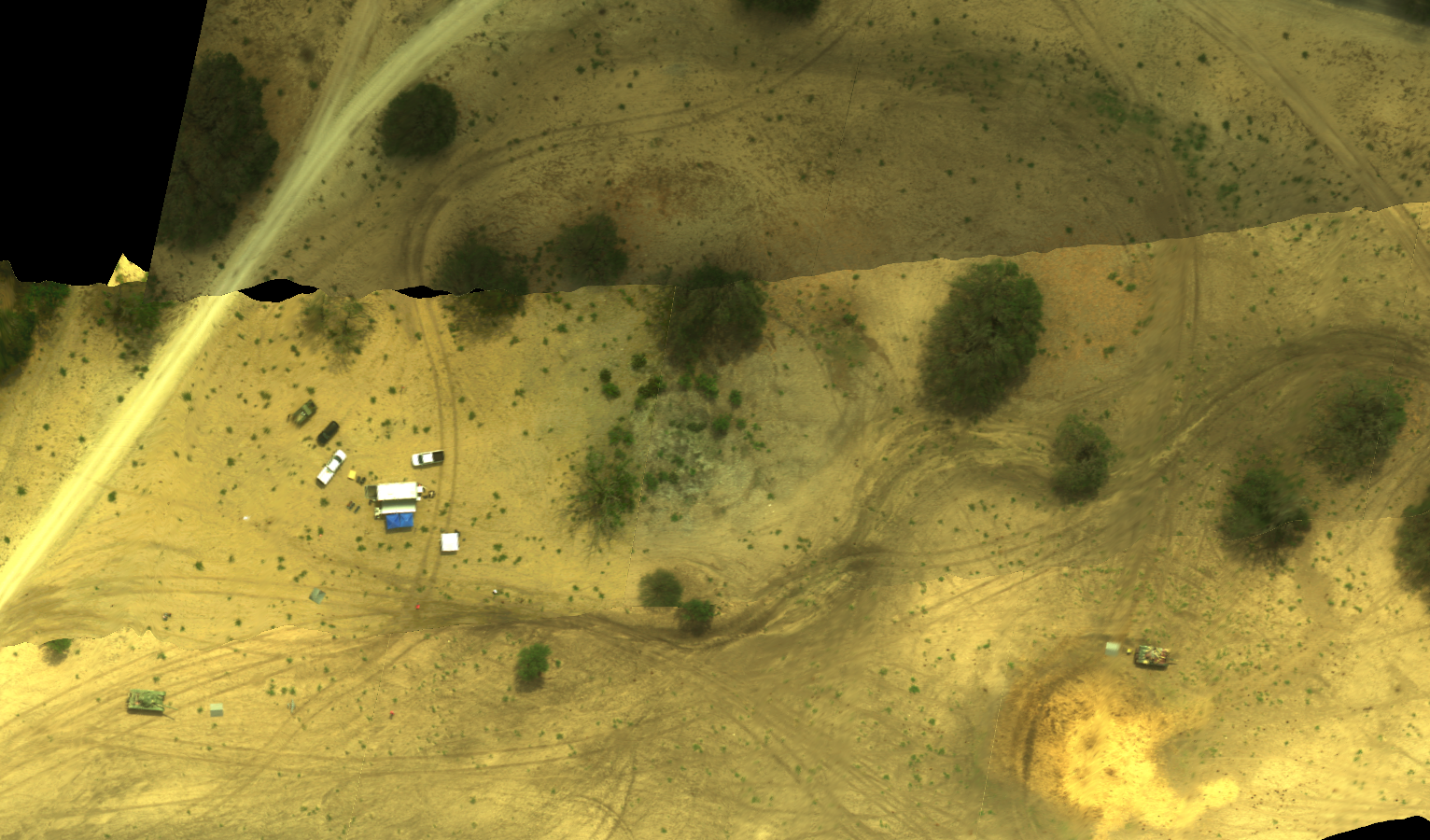}
\caption 
{ \label{fig:rgb_uvalde}
Captured Uvalde RGB image } 
\end{figure}

Figure~\ref{fig:rgb_uvalde} shows the captured RGB data of the staged area, with black areas where data was not captured. The main operations infrastructure is in the center-left of the image, and the two tanks are at the bottom corners. As this flight was done in parallel with system development, the transmission system and anomaly detector were not active at the time; the system only recorded raw data to disk. Therefore, we do not have processing time and transmission metrics available. However, the data is still useful to evaluate detection performance after the fact.

\section{Results} \label{sec:results}

After describing the data collections in Section~\ref{sec:methodology}, we now display the results, and explore the individual pieces to understand how the system is performing relative to our design. We first examine the performance timings and confirm real-time behavior. Next, we visually examine the anomaly data to confirm that it produces useful detections, then compare the detected anomalies to a ground truth to quantify detector performance. Finally, we test wireless range and reliability in some additional experiments. In Section~\ref{sec:discussion}, we discuss what we learned from the results, plus some limitations and possible improvements.

\subsection{Performance} \label{subsec:performance}

To characterize our goal of real-time performance, the system also recorded the processing time required for each step of each cube. The times are listed in Table~\ref{tab:processing} for each Memphis flight, plus times across all four flights. The table also includes a longer duration bench test where the system was not flown and the imager was not exposed to light. Processing time is not available for the Uvalde flight.

\begin{table}
\caption{Per-step cube processing time (average ± one standard deviation) \label{tab:processing}}      
\begin{tabularx}{\textwidth}{cc| cccc}
\toprule
\textbf{Scenario} & \textbf{Duration} & \textbf{Save} & \textbf{Calibrate} & \textbf{Detect} & \textbf{Transmit} \\
\midrule
\midrule

Bench Test & 20m50s & 2.81s ± 0.10s & 2.25s ± 0.13s & 1.75s ± 0.11s & 3.33s ± 0.04s \\
\midrule
Slow, No Tank & 4m13s & 2.88s ± 0.17s & 2.47s ± 0.14s & 1.79s ± 0.17s & 3.37s ± 0.07s \\
Fast, No Tank & 2m12s & 2.97s ± 0.14s & 2.49s ± 0.16s & 1.78s ± 0.13s & 3.43s ± 0.06s \\
Slow, Tank & 4m13s & 2.95s ± 0.14s & 2.47s ± 0.11s & 1.78s ± 0.10s & 3.36s ± 0.03s \\
Fast, Tank & 2m12s & 2.95s ± 0.14s & 2.58s ± 0.13s & 1.80s ± 0.13s & 3.36s ± 0.08s \\
\midrule
All Memphis & 12m52s & 2.93s ± 0.15s & 2.49s ± 0.14s & 1.78s ± 0.14s & 3.38s ± 0.07s \\

\bottomrule
\end{tabularx}
\end{table}

Acquisition of one cube always takes 4.02 seconds (1000 lines at 249Hz). The cube is passed in sequence through the previously described steps of saving its raw data to disk, performing radiometric calibration, running the RX detection algorithm, and finally packetization and transmission. In practice, the steps execute in parallel, so each works on a different cube. Therefore, the system operates in real-time as long as every step requires less time than acquisition.

As all times are under the 4.02 second requirement, the system indeed operated in real time. The bench test confirms that the system does not slow down during longer flights, and that the detection time does not significantly depend on the imaged data.

Latency is also an important parameter in a real-time system. We define the latency for our system as the time from when acquisition of a cube starts to when that cube's anomaly information is displayed. If it were acceptable for this to take ten minutes, then there would be no need to do any on-board computation for our five minute flight!

As a particular cube is processed in sequence through the steps, its latency is the sum of the time required for each step. Reception and processing in the GCS is also required before display, but that is mostly concurrent with transmission, and over a relatively small amount of data. We neglect this GCS latency as by design it only adds a few rendering frames, i.e. less than 0.1s.

Totaling acquisition time and the average time required for each step across all flights produces an average latency of 14.60 seconds. This time is well below the total flight time, reinforcing our real-time operation. The low standard deviation of our timing measurements, plus the subjective observations during our data collections, also confirm that the latency is constant and that cubes are not getting ``backed up" in processing. The bench test validates this even for longer flights.

\subsection{Detector Visual Analysis}

The (replayed) received data from all five flights is shown here without alteration. Transmission was not available during the Uvalde flight, so there is no loss from that process, but there are gaps where data was not captured due to the flight pattern. The threshold of 0.110 is consistent across all figures and was picked visually to maximize true positives and minimize false positives.

\begin{figure}
\centering
    \includegraphics[width=15.5cm]{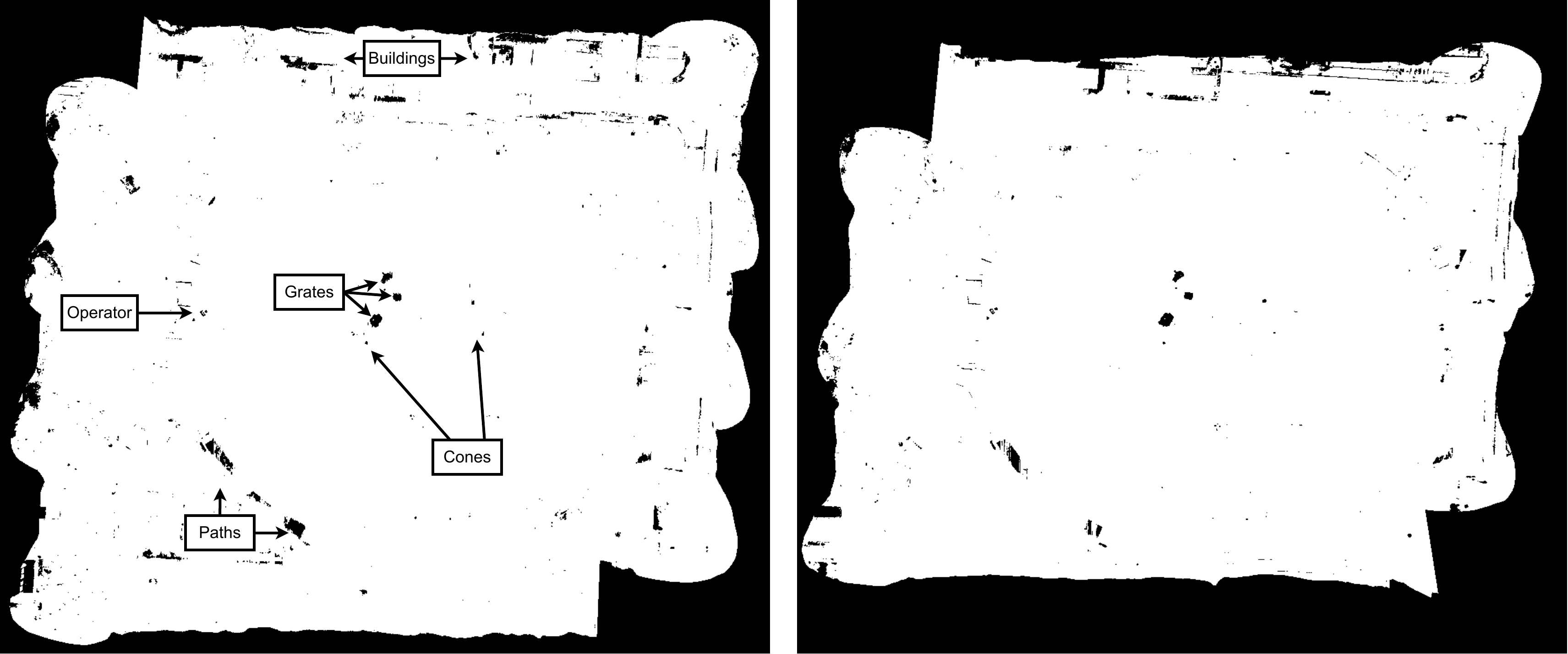}
\caption 
{ \label{fig:anomaly_notank}
Thresholded anomalies (0.110) from Memphis flights without tank (left: slow, right: fast) } 
\end{figure}

Figure~\ref{fig:anomaly_notank} shows the anomalies from the two Memphis flights without the inflatable tank. Around those, inside The Ellipse, are dotted some random traffic cones (we did not place these; not every dot is a cone). The bottom left shows some mismatched path tiles also highlighted as anomalies. There are also segments of road and buildings near the top marked as anomalous. Both flights performed similarly and highlighted the same objects.

\begin{figure}
\centering
    \includegraphics[width=15.5cm]{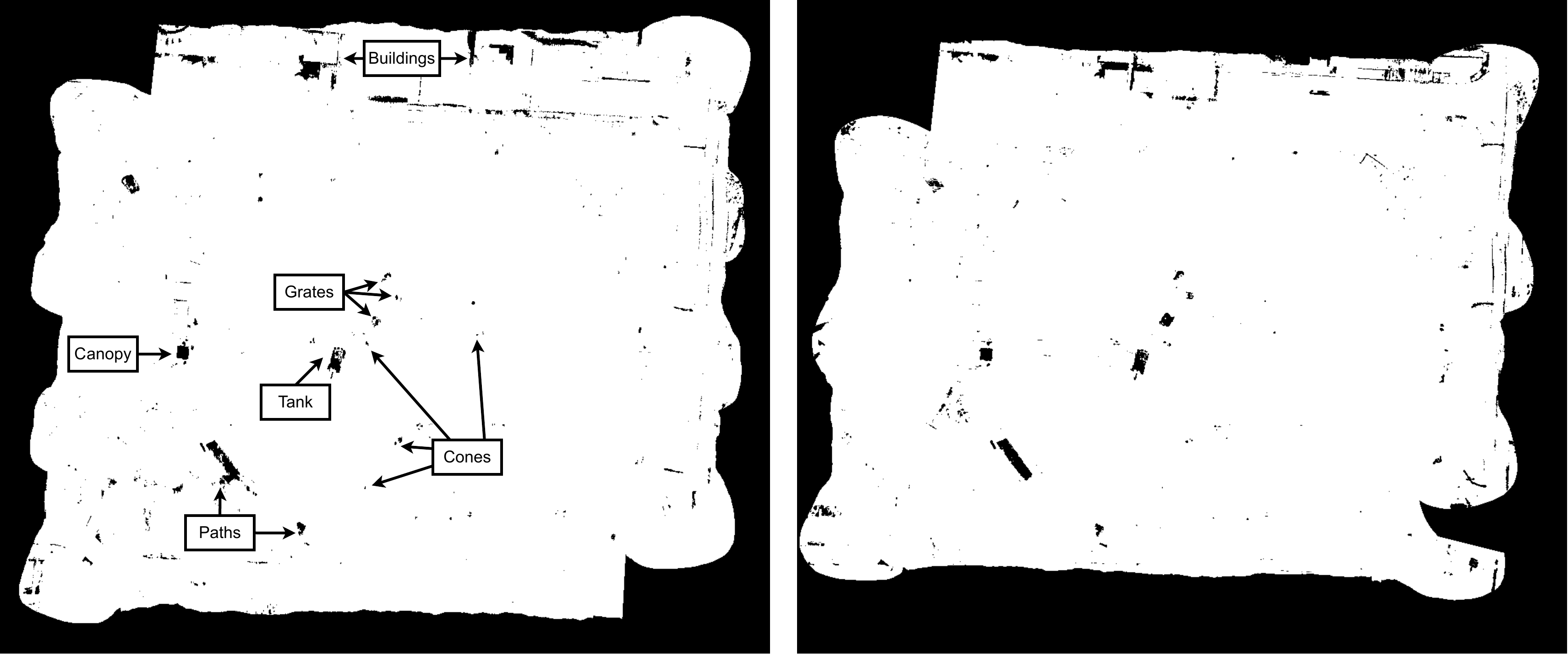}
\caption 
{ \label{fig:anomaly_tank}
Thresholded anomalies (0.110) from Memphis flights with tank (left: slow, right: fast) } 
\end{figure}

Figure~\ref{fig:anomaly_tank} shows the anomalies from the two Memphis flights with the inflatable tank. The tank and the command canopy are both clearly visible as anomalous, and most of the previous anomalies are still visible. However the prominence of the drainage grates has decreased in the slow flight, balanced by the detection of more cones.

\begin{figure}
\centering
\includegraphics[width=12cm]{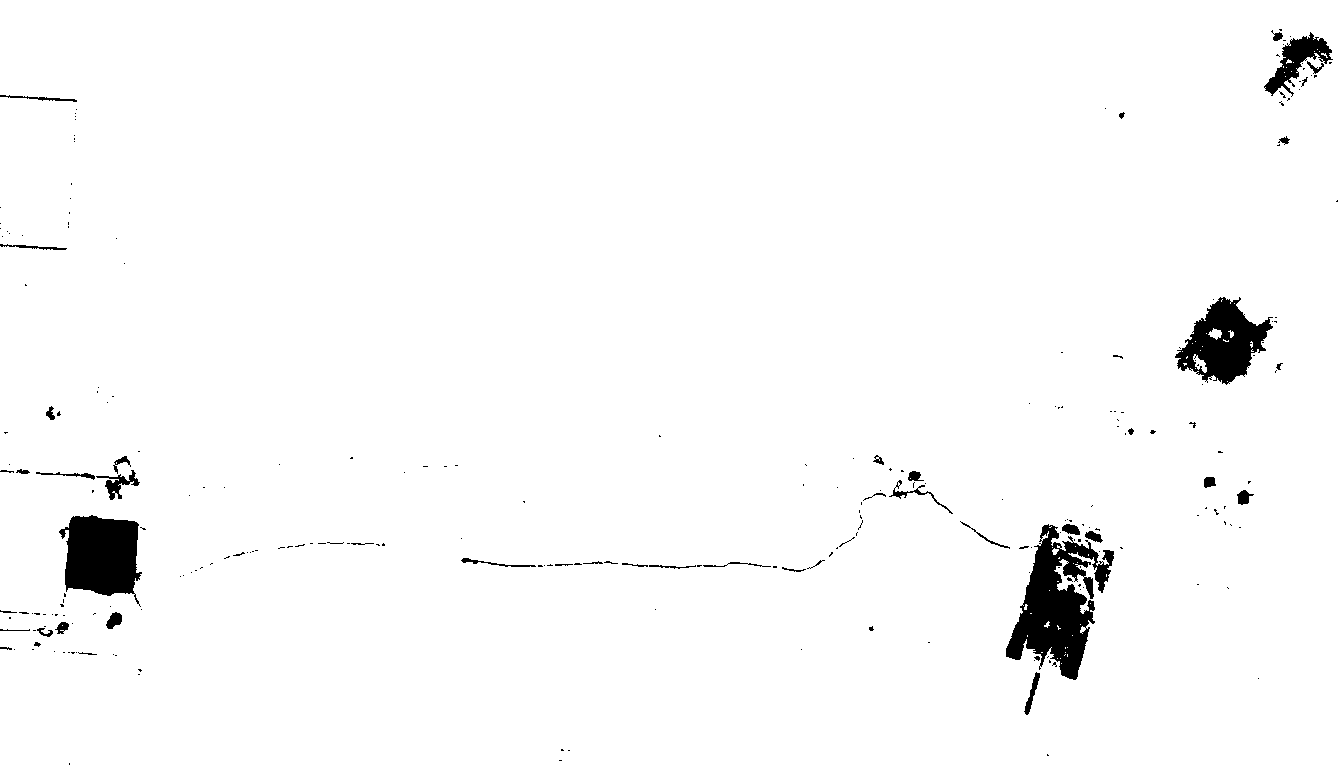}
\caption 
{ \label{fig:anomaly_tank_closeup}
Thresholded anomaly (0.110) closeup of tank on Memphis fast flight } 
\end{figure}

Figure~\ref{fig:anomaly_tank_closeup} zooms in on the data around the tank and command canopy. On the left the canopy is clearly visible. On the right are a couple drainage grates, a traffic cone, and of course the tank. The extension cord powering the tank's inflator is now visible across the middle; this was hidden earlier due to the lower resolution of the whole image. Spacecube lets the user easily zoom in on the area and examine it more closely.

\begin{figure}
\centering
\includegraphics[width=5in]{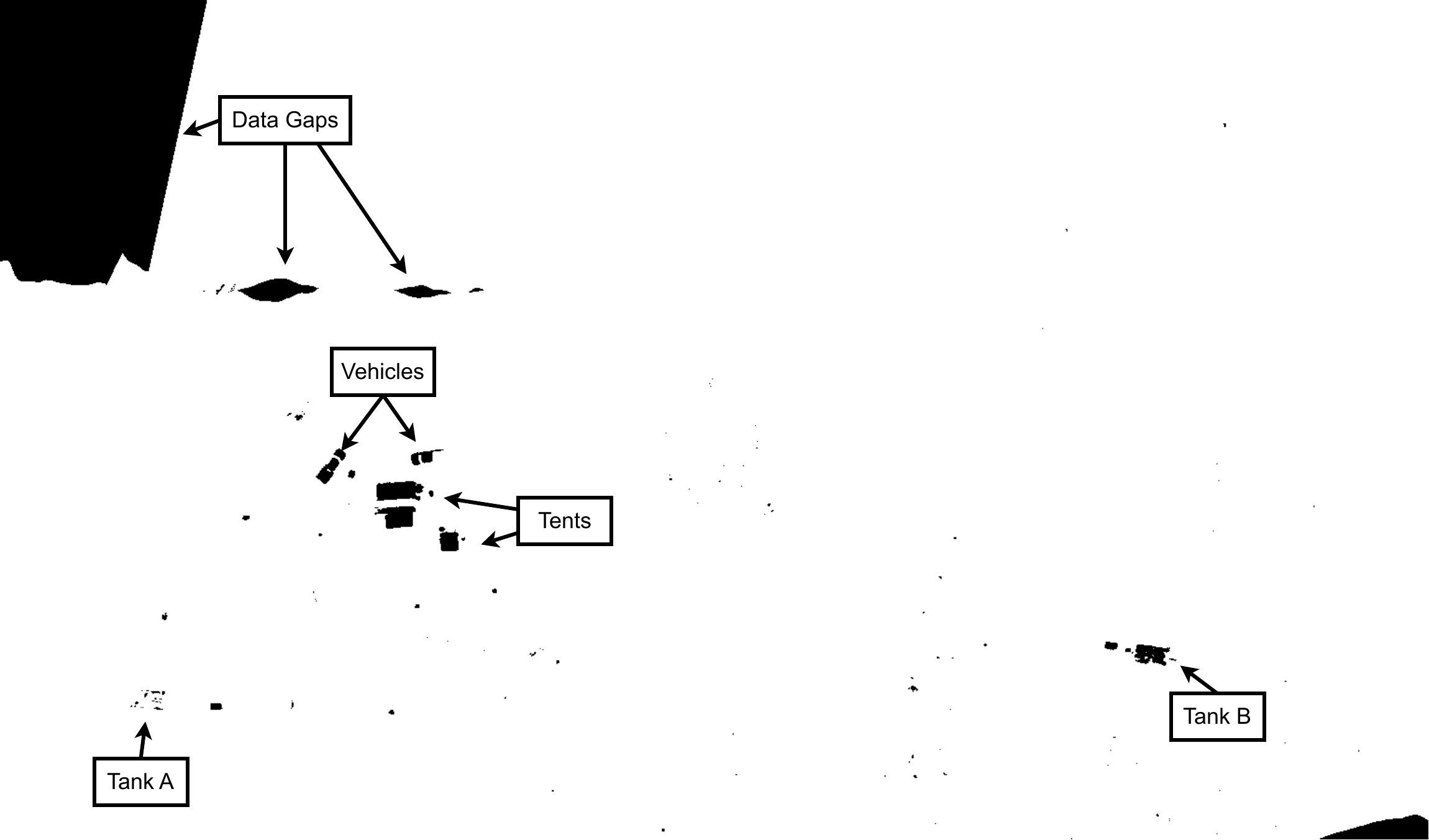}
\caption 
{ \label{fig:anomaly_uvalde}
Thresholded anomaly (0.110) overview of Uvalde } 
\end{figure}

Figure~\ref{fig:anomaly_uvalde} shows the anomalies detected in the Uvalde flight. The tanks and supporting infrastructure are clearly visible. Tank A has less anomalous pixels than tank B, but both are still visible. There are some false positives from small vegetation, but overall almost all man-made objects are easily highlighted. This shows similar results to the other flights despite the different background, scene, and objects.

\subsection{Detector Quantification}

Figure~\ref{fig:flight_results} shows the detector Receiver Operating Characteristic (ROC) curves for each flight, comparing to manually annotated ground truths. On the pre-transmission RGB data, we annotated pixels covering man-made objects as positive detections, including objects we did not place in the scene. All other pixels, including those of large man-made infrastructure like paths, roads, and buildings, were annotated negative. The curves were generated using the non-georectified anomaly scores after reception; pixels lost in transmission were assumed to have a score of zero.

\FigureSize{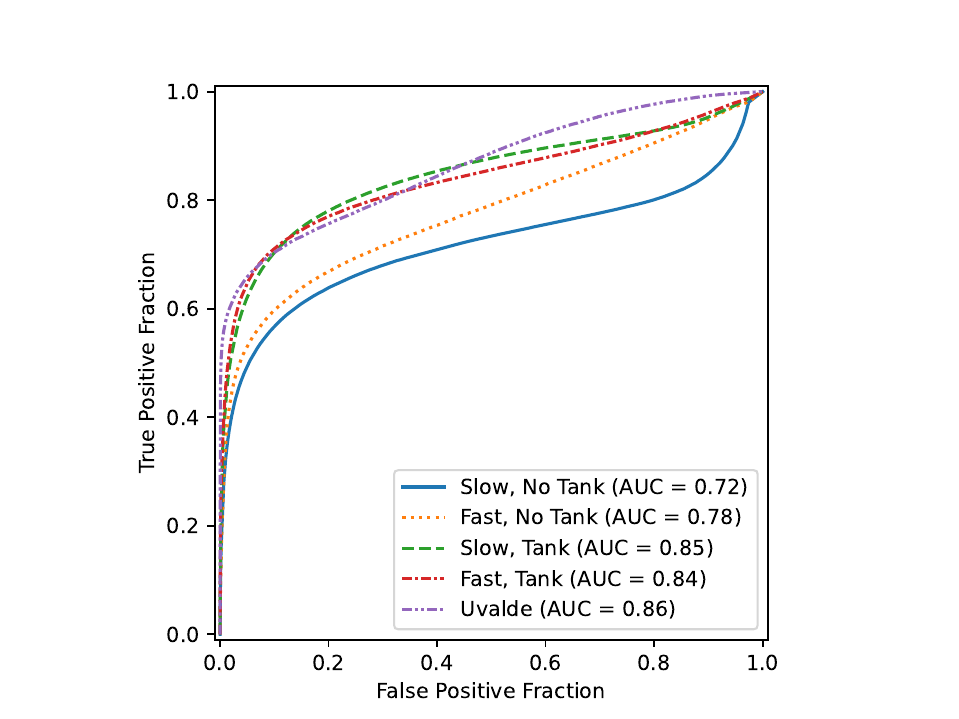}{8cm}{flight_results}{Flight Receiver Operating Characteristic (ROC) curves}

The detector performs reasonably well relative to our ground truth of man-made objects, with all AUC values similar to our previous paper results, validating that our choice of algorithm applies to our new data and scenario. The Uvalde flight also performed similar to the other flights even though the background, scene, and objects were different, confirming some generalizability across environments.

The detector does perform somewhat worse on the flights without the tank, perhaps because fewer pixels were annotated as anomalous. The flights with the tank had roughly 67\% more so annotated, as we placed several large objects in the scene but did not remove any existing ones.

Losing pixels during transmission does reduce overall accuracy slightly. The ROC curve AUC values for the lossless data are, on average, 0.97\% higher than the values for the data actually received, using our assumption that lost pixels have an anomaly score of zero. On the other hand, our 16-bit half precision encoding has negligible impact on AUC. Calculating a ROC using the raw 32 bit scores produces AUC values within 0.00013\% of the lossless results, confirming that the encoding does not significantly limit data fidelity.

\subsection{Transmission}

Analyzing the completion percentage of each cube (i.e. what fraction of lines were successfully received after error correction), the packet reception rate for the four Memphis flights had a mean of 97.91\% and a standard deviation of 5.28\% over a maximum reception distance of 200m; 71\% of cubes were received with no missing lines. Our radio link worked well, though it was limited in practice by operator skill in holding the receiver as poor orientation degrades reception. Repeating the same flights with a different operator showed rates exceeding 99.84\%, suggesting antennas on the back of the GCS may not be ideal.

We also performed two reception range tests in more rural areas of Tennessee. The primary objective of these tests was capture of particular creek and stream features for partner projects. Therefore, they were performed without an experienced operator and with some confounding factors such as trees, and sight lines close to the horizon. These tests were not used to evaluate anomaly detection  performance.

The first test used a flight path 6.4km long and 80m above ground, a maximum reception distance of 2.0km, and positioned the receiver away from trees with good sight to the UAS. The total flight collection was 234 cubes with a reception rate of 86.63\%. We further analyzed by dividing the flight pattern into ``favorable" and ``unfavorable" portions, with the latter similar to the former but with a 180 degree heading offset, and both reaching to 2.0km. The favorable portion (2 legs, 107 cubes) had a reception rate of 96.38\%, while the unfavorable portion (2 legs, 127 cubes) had a rate of 78.41\%. During the favorable portion, most cubes even at the farthest distances were 100\% received. This shows a significant difference in the effect of antenna positioning and orientation on the UAS, and suggests effort should be spent optimizing it.

The second test used a flight path 6.1km long and 80m above ground with a maximum distance of 2.5km, but was limited to putting the receiver in a location with some trees limiting sight to the UAS. The total flight collection was 226 cubes with a reception rate of 64.42\%. A favorable portion (1 leg, 1.7km distance, 83 cubes) showed a reception rate of 99.37\%, while the unfavorable portion (2 legs, 2.5km distance, 143 cubes) showed a reception rate of only 44.13\%. Some contiguous data was received from the farthest distances, but the overall rate there was very poor. The favorable portion affirms good performance at long distances, but the unfavorable portion emphasizes the harm a poor environment can cause.

\section{Discussion} \label{sec:discussion}

Having presented the results and related measurements in Section~\ref{sec:results}, we now discuss some findings, explanations, and implications based on that information. We additionally explain some known and observed limitations, along with possible improvements.

\subsection{Performance}

Our results show real-time performance, and the latency means data is available on the ground in under 15 seconds after flying over an object. Therefore, operator analysis can start during the UAS takeoff sequence, or even before. This is a novel capability in hyperspectral anomaly detection systems, and it opens up new applications for the data.

However, the air SBC is close to CPU capacity as suggested by the step timings in Table~\ref{tab:processing}. Our design of each step processing a separate cube requires several cubes to be in memory at once, pushing the SBC near RAM capacity as well. Therefore, adding pre-processing or a more sophisticated detector is likely to run over budget, and a more powerful SBC would be required. We do not yet use its integrated GPU, which would likely speed up the calibration and detection steps. There are more powerful SBCs in the UP Squared series, among many possibilities, but others have not yet been tested.

The table also shows that calibration is slightly faster on the bench. This is likely because the scene and thus camera settings were constant, reducing overhead from calibrating with many per-setting tables. Saving is also slightly faster as the static dark scene compresses more efficiently, reducing the data to save.

Regarding data latency, the air system computation structure was chosen for ease of implementation rather than maximum efficiency or minimum latency. Each step of processing a cube adds several seconds of latency as the steps are performed on a particular cube in sequence. Multiple steps could run on the same cube in parallel, and steps related to transmission could be prioritized, perhaps halving latency.

Switching to a line-based processing architecture would massively reduce latency, and save RAM as well. Most steps currently perform the same process in a loop over the 1000 lines in a cube. If that process was performed only on one line, the next step could start on that line after 1000x less time, and entire cubes wouldn't need to stay in memory. However, this requires much tighter programming integration among the steps and could amplify overhead from passing data between them. Even with the current detector that examines 1000 lines at a time, this would reduce latency to under five seconds (acquisition of 1000 lines, plus a bit). A line-based detector could reduce latency to well under one second, enhancing the user experience and real-time value.

\subsection{Detection}

Though based on a straightforward implementation of the classic RX architecture, our detector performed well in our experiments and was able to support real-time operation. Overall the thresholded result matches expectations and easily highlights unique objects in the scene for an operator. The quantitative performance, even using our conservative anomaly definition of ``man-made object", also aligns with previous results. The RX detector requires no tuning or configuration, and so it is immediately ready to produce results for any potential scene.

There is some difference in behavior between our tests and compared to our ground truth. In the slow flight with the tank, the large influence on the distribution from the relatively long period flying over the tank reduced the prominence of the nearby drainage grates. The detector also considered some visually distinct parts of buildings and paths as anomalous, which makes sense in the abstract but disagreed with our ground truth.

The detector also seems to be insensitive to dark objects, independent of cloud cover or lighting. In the Uvalde data, the black vehicle is completely missed, and black equipment in the Memphis data is also overlooked. Shadows are not generally picked up either. This discrepancy may be due to the fact that there is more ``room" for an object to be far from the distribution if it is brighter, as an object cannot be darker than zero. This likely needs a more sophisticated detector to address.

The detector is only as good as the imager, and there are some spectral anomalies that can only be seen in the UV or longer wavelength IR bands, making our VNIR imager unsuitable to detect those. Our architecture is compatible with any line-scan hyperspectral imager of any number or range of bands or samples. However, additional programming work would be required to integrate a different driver for such an imager, and a higher resolution imager would need more processing power.

There are other limitations in the acquisition system which can negatively influence the detector. The biggest limitation is the lack of reflectance calibration, which does not matter for this particular detector, but limits the choice of algorithms to linear ones. As discussed in Section~\ref{subsec:arch_acq}, reflectance calibration requires additional inputs our system does not currently have. The imager is also noisier in low light, making operation less reliable early or late in the day when the sun is low in the sky, or in poor weather.

Related to weather, the detector is impacted by variable cloud cover, where different areas on the ground within a cube may be within cloud shadow or sunlight. This creates a wide distribution and so marks less objects as anomalous. As the distribution is recomputed for each cube, a cube which is entirely in cloud shadow or entirely in sunlight is not affected. An accurate reflectance calibration could compensate for this effect, but making the calibration adaptive with variable shade would require at least an additional sensor.

\subsection{Transmission}

The transmission system works very well in good conditions and is able to transmit cubes without loss from long range. However, especially in our additional experiments, ``good conditions" covered fewer portions of flights than was ideal. The system needs some work on both transmitter and receiver to address radio performance and antenna positioning to improve overall effectiveness.

Our transmission strategy of sending lines and FEC groups in sequence could be made more robust as well. If there is a moderate communication gap (e.g. 50 lines, ~0.2s), the current group will be damaged beyond repair and a large black region will appear to the operator. Sending multiple groups in parallel would cause a gap to damage a small part of several groups, which is repairable. Shuffling lines and samples would cause any damage to be spread out among the cube, shrinking the size of each black region and reducing the risk of missing an anomalous object. However, both strategies would increase latency.

The impact of data lost in transmission is visible both in the displayed RGB data and quantitatively in the (slightly) reduced detector performance figures. Our strategy of coloring missed areas black and assuming zero anomaly score is easiest and most conservative, but it is also most harmful to operator experience and system performance. Interpolating missed pixels from data nearby would improve these aspects and avoid introducing known-incorrect anomaly and RGB information that may distract an operator or algorithm.

\section{Conclusions}

We have described and demonstrated, to our knowledge, the first system to perform real-time aerial anomaly detection on a UAS, including a Resonon Pika L hyperspectral sensor, an UP Squared Pro SBC running the RX anomaly detection algorithm, an inexpensive and high-performance radio link based on wfb-ng and consumer WiFi cards, and a Valve Steam Deck running our Spacecube software to georectify and display the results.

Our earlier results studying RX algorithm variants on the UP Squared Pro SBC informed us on a good strategy and parameters to use. The SBC is sufficiently powerful for a spectrally binned global RX algorithm, and this algorithm provides qualitatively useful detection results that also match well to ground truth. We confirmed that advanced variants were not necessary to achieve good real-time performance and detection results.

Our earlier work on Spacecube provided the backbone of our ground control station experience. Spacecube allows the operator to zoom, pan, tweak georectification for collection non-idealities, and switch viewing modes. Switching modes and adjusting georectification takes under one second; all other inputs are processed instantly.

The radios are inexpensive, powerful, and work very well in practice, though the system can be adapted to any radios which provide sufficient continuous bandwidth. In our implementation, the error correction and allocation of bandwidth provide a high quality of service when conditions permit, with bounded latency. As in all radio systems, ultimate performance is limited by the environment and antennas.

In the future, operating the completed system on a fixed-wing UAS would make it easier to conduct longer flights to search for more anomalies and better determine the limits of the radio range. Implementing different algorithms in the air and viewing modes on the ground could extend the system's application, particularly after studying how operators use the system. Finally, the georectified data could be fed into a machine learning system to further filter the data and/or autonomously direct the UAS to more thoroughly investigate anomalous areas.

Our system implementation demonstrates for the first time that real-time anomaly detection is possible and useful with UAS-practical computational power. Spacecube's innovative real-time georectification facilitates interactive investigation of the results on the ground. Overall, performing detection in the air allows a useful data summary to be transmitted in real time for immediate operator analysis, while still leveraging the full spectrum of information available from a hyperspectral imager.

\section*{Acknowledgements}

This material is based upon work supported by the U.S. Army Contracting Command under Contract No. W911QX-24-C-0012.  Any opinions, findings and conclusions or recommendations expressed in this material are those of the author(s) and do not necessarily reflect the views of the Army Contracting Command.

We are very grateful to Hidalgo et al.~\cite{hidalgo_efficient_2021} and the Spectral Python authors for sharing their source code publicly. We are also grateful to our lab members and partner universities for supporting field operations and data collections.

The authors declare no conflicts of interest. The funders participated in the study by reviewing and approving the manuscript for public release.

The authors did not use Large Language Models (LLMs) for any component of this paper, including manuscript authoring and internal review, data collection and analysis, and development of the associated software.

The source code for the system demonstrated in this paper is available under a GPLv3 (or later) license at GitHub~\cite{watson_spacecube_2025}. The Memphis data is available at Zenodo~\cite{watson_hyperspectral_2025} under a CC-BY-4.0 license. Due to contractual limitations, the Uvalde data is not publicly available; contact the authors for more information.

\bibliographystyle{IEEEtran}
\bibliography{IEEEabrv,report}

@article{he_recursive_2023,
	title = {Recursive {RX} with {Extended} {Multi}-{Attribute} {Profiles} for {Hyperspectral} {Anomaly} {Detection}},
	volume = {15},
	issn = {2072-4292},
	url = {https://www.mdpi.com/2072-4292/15/3/589},
	doi = {10.3390/rs15030589},
	language = {en},
	number = {3},
	urldate = {2023-02-20},
	journal = {Remote Sensing},
	author = {He, Fang and Yan, Shuai and Ding, Yao and Sun, Zhensheng and Zhao, Jianwei and Hu, Haojie and Zhu, Yujie},
	month = jan,
	year = {2023},
	pages = {589},
}

@article{rossi_rx_2014,
	title = {{RX} architectures for real-time anomaly detection in hyperspectral images},
	volume = {9},
	issn = {1861-8200, 1861-8219},
	url = {http://link.springer.com/10.1007/s11554-012-0292-3},
	doi = {10.1007/s11554-012-0292-3},
	language = {en},
	number = {3},
	urldate = {2023-02-20},
	journal = {Journal of Real-Time Image Processing},
	author = {Rossi, A. and Acito, N. and Diani, M. and Corsini, G.},
	month = sep,
	year = {2014},
	pages = {503--517},
}

@article{garzon_anomaly_2012,
	title = {Anomaly detection based on a parallel kernel {RX} algorithm for multicore platforms},
	volume = {6},
	issn = {1931-3195},
	url = {http://remotesensing.spiedigitallibrary.org/article.aspx?doi=10.1117/1.JRS.6.061503},
	doi = {10.1117/1.JRS.6.061503},
	language = {en},
	number = {1},
	urldate = {2023-02-20},
	journal = {Journal of Applied Remote Sensing},
	author = {Garzón, Ester M.},
	month = may,
	year = {2012},
	pages = {061503},
}

@article{zhao_fast_2018,
	title = {Fast {Real}-{Time} {Kernel} {RX} {Algorithm} {Based} on {Cholesky} {Decomposition}},
	volume = {15},
	issn = {1545-598X, 1558-0571},
	url = {https://ieeexplore.ieee.org/document/8432482/},
	doi = {10.1109/LGRS.2018.2859426},
	language = {en},
	number = {11},
	urldate = {2023-02-20},
	journal = {IEEE Geoscience and Remote Sensing Letters},
	author = {Zhao, Chunhui and Xi-Feng, Yao},
	month = nov,
	year = {2018},
	pages = {1760--1764},
}

@article{kwon_kernel_2005,
	title = {Kernel {RX}-algorithm: a nonlinear anomaly detector for hyperspectral imagery},
	volume = {43},
	issn = {0196-2892},
	shorttitle = {Kernel {RX}-algorithm},
	url = {http://ieeexplore.ieee.org/document/1386510/},
	doi = {10.1109/TGRS.2004.841487},
	language = {en},
	number = {2},
	urldate = {2023-02-20},
	journal = {IEEE Transactions on Geoscience and Remote Sensing},
	author = {Kwon, Heesung and Nasrabadi, N.M.},
	month = feb,
	year = {2005},
	pages = {388--397},
}

@article{hidalgo_efficient_2021,
	title = {Efficient {Nonlinear} {RX} {Anomaly} {Detectors}},
	volume = {18},
	issn = {1545-598X, 1558-0571},
	url = {http://arxiv.org/abs/2012.05799},
	doi = {10.1109/LGRS.2020.2970582},
	language = {en},
	number = {2},
	urldate = {2023-02-20},
	journal = {IEEE Geoscience and Remote Sensing Letters},
	author = {Hidalgo, José A. Padrón and Pérez-Suay, Adrián and Nar, Fatih and Camps-Valls, Gustau},
	month = feb,
	year = {2021},
	pages = {231--235},
}

@article{reed_adaptive_1990,
	title = {Adaptive multiple-band {CFAR} detection of an optical pattern with unknown spectral distribution},
	volume = {38},
	issn = {00963518},
	url = {http://ieeexplore.ieee.org/document/60107/},
	doi = {10.1109/29.60107},
	language = {en},
	number = {10},
	urldate = {2023-02-20},
	journal = {IEEE Transactions on Acoustics, Speech, and Signal Processing},
	author = {Reed, I.S. and Yu, X.},
	month = oct,
	year = {1990},
	pages = {1760--1770},
}

@article{lu_recent_2020,
	title = {Recent {Advances} of {Hyperspectral} {Imaging} {Technology} and {Applications} in {Agriculture}},
	volume = {12},
	issn = {2072-4292},
	url = {https://www.mdpi.com/2072-4292/12/16/2659},
	doi = {10.3390/rs12162659},
	language = {en},
	number = {16},
	urldate = {2023-02-27},
	journal = {Remote Sensing},
	author = {Lu, Bing and Dao, Phuong and Liu, Jiangui and He, Yuhong and Shang, Jiali},
	month = aug,
	year = {2020},
	pages = {2659},
}

@article{stuart_hyperspectral_2019,
	title = {Hyperspectral {Imaging} in {Environmental} {Monitoring}: {A} {Review} of {Recent} {Developments} and {Technological} {Advances} in {Compact} {Field} {Deployable} {Systems}},
	volume = {19},
	issn = {1424-8220},
	shorttitle = {Hyperspectral {Imaging} in {Environmental} {Monitoring}},
	url = {https://www.mdpi.com/1424-8220/19/14/3071},
	doi = {10.3390/s19143071},
	language = {en},
	number = {14},
	urldate = {2023-02-27},
	journal = {Sensors},
	author = {Stuart, Mary B. and McGonigle, Andrew J. S. and Willmott, Jon R.},
	month = jul,
	year = {2019},
	pages = {3071},
}

@article{shimoni_hypersectral_2019,
	title = {Hypersectral {Imaging} for {Military} and {Security} {Applications}: {Combining} {Myriad} {Processing} and {Sensing} {Techniques}},
	volume = {7},
	issn = {2168-6831, 2473-2397, 2373-7468},
	shorttitle = {Hypersectral {Imaging} for {Military} and {Security} {Applications}},
	url = {https://ieeexplore.ieee.org/document/8738016/},
	doi = {10.1109/MGRS.2019.2902525},
	number = {2},
	urldate = {2023-02-27},
	journal = {IEEE Geoscience and Remote Sensing Magazine},
	author = {Shimoni, Michal and Haelterman, Rob and Perneel, Christiaan},
	month = jun,
	year = {2019},
	pages = {101--117},
}

@misc{resonon_resonon_2023,
	title = {Resonon {Pika} {L} {Datasheet}},
	url = {https://resonon.com/content/files/Resonon---Camera-Data-Sheets-Pika-L.pdf},
	urldate = {2025-11-13},
	publisher = {Resonon},
	author = {{Resonon}},
	year = {2023},
}

@misc{raspberry_pi_raspberry_2023,
	title = {Raspberry {Pi} 4 {Model} {B} {Datasheet}},
	url = {https://datasheets.raspberrypi.com/rpi4/raspberry-pi-4-datasheet.pdf},
	publisher = {Raspberry Pi},
	author = {{Raspberry Pi}},
	year = {2023},
}

@misc{nvidia_embedded_2023,
	title = {Embedded {Systems} with {Jetson}},
	url = {https://www.nvidia.com/en-us/autonomous-machines/embedded-systems/},
	author = {{NVIDIA}},
	year = {2023},
}

@misc{up_up_2023,
	title = {{UP} {Squared} {Pro} {Datasheet}},
	url = {https://up-board.org/wp-content/uploads/2020/12/UP_Squared_Pro_Datasheet_v3.pdf},
	author = {{UP}},
	year = {2023},
}

@article{fawcett_introduction_2006,
	title = {An introduction to {ROC} analysis},
	volume = {27},
	issn = {01678655},
	url = {https://linkinghub.elsevier.com/retrieve/pii/S016786550500303X},
	doi = {10.1016/j.patrec.2005.10.010},
	language = {en},
	number = {8},
	urldate = {2023-02-27},
	journal = {Pattern Recognition Letters},
	author = {Fawcett, Tom},
	month = jun,
	year = {2006},
	pages = {861--874},
}

@article{angel_automated_2019,
	title = {Automated georectification and mosaicking of {UAV}-based hyperspectral imagery from push-broom sensors},
	volume = {12},
	copyright = {https://creativecommons.org/licenses/by/4.0/},
	issn = {2072-4292},
	url = {https://www.mdpi.com/2072-4292/12/1/34},
	doi = {10.3390/rs12010034},
	language = {en},
	number = {1},
	urldate = {2024-10-07},
	journal = {Remote Sensing},
	author = {Angel, Yoseline and Turner, Darren and Parkes, Stephen and Malbeteau, Yoann and Lucieer, Arko and McCabe, Matthew F.},
	month = dec,
	year = {2019},
	pages = {34},
}

@article{warren_data_2014,
	title = {Data processing of remotely sensed airborne hyperspectral data using the {Airborne} {Processing} {Library} ({APL}): geocorrection algorithm descriptions and spatial accuracy assessment},
	volume = {64},
	issn = {00983004},
	shorttitle = {Data processing of remotely sensed airborne hyperspectral data using the airborne processing library ({APL})},
	url = {https://linkinghub.elsevier.com/retrieve/pii/S0098300413002938},
	doi = {10.1016/j.cageo.2013.11.006},
	language = {en},
	urldate = {2024-10-07},
	journal = {Computers \& Geosciences},
	author = {Warren, Mark A. and Taylor, Benjamin H. and Grant, Michael G. and Shutler, Jamie D.},
	month = mar,
	year = {2014},
	pages = {24--34},
}

@article{k_c_lawrence_calibration_2003,
	title = {Calibration of a pushbroom hyperspectral imaging system for agricultural inspection},
	volume = {46},
	issn = {2151-0059},
	url = {http://elibrary.asabe.org/abstract.asp??JID=3&AID=12940&CID=t2003&v=46&i=2&T=1},
	doi = {10.13031/2013.12940},
	language = {en},
	number = {2},
	urldate = {2024-12-02},
	journal = {Transactions of the ASAE},
	author = {{K. C. Lawrence} and {B. Park} and {W. R. Windham} and {C. Mao}},
	year = {2003},
}

@article{geladi_hyperspectral_2004,
	title = {Hyperspectral imaging: calibration problems and solutions},
	volume = {72},
	copyright = {https://www.elsevier.com/tdm/userlicense/1.0/},
	issn = {01697439},
	shorttitle = {Hyperspectral imaging},
	url = {https://linkinghub.elsevier.com/retrieve/pii/S0169743904000371},
	doi = {10.1016/j.chemolab.2004.01.023},
	language = {en},
	number = {2},
	urldate = {2024-12-02},
	journal = {Chemometrics and Intelligent Laboratory Systems},
	author = {Geladi, Paul and Burger, Jim and Lestander, Torbjörn},
	month = jul,
	year = {2004},
	pages = {209--217},
}

@article{yang_ccd_2003,
	title = {A {CCD} camera‐based hyperspectral imaging system for stationary and airborne applications},
	volume = {18},
	issn = {1010-6049, 1752-0762},
	url = {http://www.tandfonline.com/doi/abs/10.1080/10106040308542274},
	doi = {10.1080/10106040308542274},
	language = {en},
	number = {2},
	urldate = {2024-12-03},
	journal = {Geocarto International},
	author = {Yang, Chenghai and Everitt, James H. and Davis, Michael R. and Mao, Chengye},
	month = jun,
	year = {2003},
	pages = {71--80},
}

@inproceedings{watson_evaluation_2023,
	address = {Orlando, United States},
	title = {Evaluation of aerial real-time {RX} anomaly detection},
	isbn = {978-1-5106-6152-3 978-1-5106-6153-0},
	url = {https://www.spiedigitallibrary.org/conference-proceedings-of-spie/12519/2663904/Evaluation-of-aerial-real-time-RX-anomaly-detection/10.1117/12.2663904.full},
	doi = {10.1117/12.2663904},
	urldate = {2025-04-09},
	booktitle = {Algorithms, {Technologies}, and {Applications} for {Multispectral} and {Hyperspectral} {Imaging} {XXIX}},
	publisher = {SPIE},
	author = {Watson, Thomas P. and McKenzie, Kevin and Robinson, Aaron L. and Renshaw, Kyle and Driggers, Ron and Jacobs, Eddie L. and Conroy, Joseph},
	editor = {Messinger, David W. and Velez-Reyes, Miguel},
	month = jun,
	year = {2023},
	pages = {40},
}

@article{garske_erx_2025,
	title = {{ERX}: {A} {Fast} {Real}-{Time} {Anomaly} {Detection} {Algorithm} for {Hyperspectral} {Line} {Scanning}},
	volume = {63},
	copyright = {https://creativecommons.org/licenses/by/4.0/legalcode},
	issn = {0196-2892, 1558-0644},
	shorttitle = {{ERX}},
	url = {https://ieeexplore.ieee.org/document/10847782/},
	doi = {10.1109/TGRS.2025.3532225},
	urldate = {2025-08-20},
	journal = {IEEE Transactions on Geoscience and Remote Sensing},
	author = {Garske, Samuel and Evans, Bradley and Artlett, Christopher and Wong, K. C.},
	year = {2025},
	pages = {1--17},
}

@inproceedings{loke_next-gen_2025,
	address = {Orlando, United States},
	title = {Next-gen {UAV} hyperspectral processing: transforming {UAV} platforms into real-time end-user solutions with {3D} hypermesh creation},
	isbn = {978-1-5106-8699-1 978-1-5106-8700-4},
	shorttitle = {Next-gen {UAV} hyperspectral processing},
	url = {https://www.spiedigitallibrary.org/conference-proceedings-of-spie/13455/3054170/Next-gen-UAV-hyperspectral-processing--transforming-UAV-platforms-into/10.1117/12.3054170.full},
	doi = {10.1117/12.3054170},
	language = {en},
	urldate = {2025-11-10},
	booktitle = {Algorithms, {Technologies}, and {Applications} for {Multispectral} and {Hyperspectral} {Imaging} {XXXI}},
	publisher = {SPIE},
	author = {Løke, Trond and Sivertsen, Agnar and Stødle, Daniel and Løkse, Sigurd and Bohman, Axel and Schläfpher, Daniel},
	editor = {Messinger, David W. and Velez-Reyes, Miguel},
	month = may,
	year = {2025},
	pages = {9},
}

@misc{ardupilot_project_sik_2025,
	title = {{SiK} {Telemetry} {Radio}},
	url = {https://ardupilot.org/copter/docs/common-sik-telemetry-radio.html},
	urldate = {2025-11-10},
	publisher = {ArduPilot},
	author = {{ArduPilot Project}},
	year = {2025},
}

@inproceedings{fabra_methodology_2017,
	address = {Las Vegas, NV, USA},
	title = {A methodology for measuring {UAV}-to-{UAV} communications performance},
	isbn = {978-1-5090-6196-9},
	url = {http://ieeexplore.ieee.org/document/7983120/},
	doi = {10.1109/CCNC.2017.7983120},
	language = {en},
	urldate = {2025-11-10},
	booktitle = {2017 14th {IEEE} {Annual} {Consumer} {Communications} \& {Networking} {Conference} ({CCNC})},
	publisher = {IEEE},
	author = {Fabra, Francisco and Calafate, Carlos T. and Cano, Juan Carlos and Manzoni, Pietro},
	month = jan,
	year = {2017},
	pages = {280--286},
}

@inproceedings{asadpour_ground_2013,
	address = {Miami Florida USA},
	title = {From ground to aerial communication: dissecting {WLAN} 802.11n for the drones},
	isbn = {978-1-4503-2364-2},
	shorttitle = {From ground to aerial communication},
	url = {https://dl.acm.org/doi/10.1145/2505469.2505472},
	doi = {10.1145/2505469.2505472},
	language = {en},
	urldate = {2025-11-10},
	booktitle = {Proceedings of the 8th {ACM} international workshop on {Wireless} network testbeds, experimental evaluation \& characterization},
	publisher = {ACM},
	author = {Asadpour, Mahdi and Giustiniano, Domenico and Hummel, Karin Anna},
	month = sep,
	year = {2013},
	pages = {25--32},
}

@misc{doodle_labs_mini-oem_2024,
	title = {mini-{OEM} {Mesh} {Rider} {Radio}},
	url = {https://techlibrary.doodlelabs.com/doodle-labs-mini-oem-mesh-rider-radio-24002482-mhz},
	urldate = {2025-11-10},
	publisher = {Doodle Labs},
	author = {{Doodle Labs}},
	year = {2024},
}

@inproceedings{gunther_analysis_2014,
	address = {Krakow, Poland},
	title = {Analysis of injection capabilities and media access of {IEEE} 802.11 hardware in monitor mode},
	isbn = {978-1-4799-0913-1},
	url = {http://ieeexplore.ieee.org/document/6838262/},
	doi = {10.1109/NOMS.2014.6838262},
	language = {en},
	urldate = {2025-11-10},
	booktitle = {2014 {IEEE} {Network} {Operations} and {Management} {Symposium} ({NOMS})},
	publisher = {IEEE},
	author = {Gunther, Stephan M. and Leclaire, Maurice and Michaelis, Julius and Carle, Georg},
	month = may,
	year = {2014},
	pages = {1--9},
}

@inproceedings{vanhoef_testing_2023,
	address = {Guildford United Kingdom},
	title = {Testing and {Improving} the {Correctness} of {Wi}-{Fi} {Frame} {Injection}},
	isbn = {978-1-4503-9859-6},
	url = {https://dl.acm.org/doi/10.1145/3558482.3581779},
	doi = {10.1145/3558482.3581779},
	language = {en},
	urldate = {2025-11-10},
	booktitle = {Proceedings of the 16th {ACM} {Conference} on {Security} and {Privacy} in {Wireless} and {Mobile} {Networks}},
	publisher = {ACM},
	author = {Vanhoef, Mathy and Jiao, Xianjun and Liu, Wei and Moerman, Ingrid},
	month = may,
	year = {2023},
	pages = {287--292},
}

@book{wicker_reed-solomon_1999,
	title = {Reed-{Solomon} codes and their applications},
	publisher = {John Wiley \& Sons},
	author = {Wicker, Stephen B and Bhargava, Vijay K},
	year = {1999},
}

@misc{evseenko_wfb-ng_2023,
	title = {wfb-ng},
	url = {https://github.com/svpcom/wfb-ng},
	urldate = {2025-11-10},
	author = {Evseenko, Vasily},
	year = {2023},
}

@inproceedings{watson_vidpak_2025,
	address = {Orlando, United States},
	title = {Vidpak: high speed lossless scientific video compression},
	isbn = {978-1-5106-8705-9 978-1-5106-8706-6},
	shorttitle = {Vidpak},
	url = {https://www.spiedigitallibrary.org/conference-proceedings-of-spie/13458/3053812/Vidpak-high-speed-lossless-scientific-video-compression/10.1117/12.3053812.full},
	doi = {10.1117/12.3053812},
	urldate = {2025-11-10},
	booktitle = {Real-{Time} {Image} {Processing} and {Deep} {Learning} 2025},
	publisher = {SPIE},
	author = {Watson, Thomas P. and Renshaw, Kyle and Jacobs, Eddie},
	editor = {Kehtarnavaz, Nasser and Shirvaikar, Mukul V.},
	month = may,
	year = {2025},
	pages = {16},
}

@misc{watson_hyperspectral_2025,
	address = {Zenodo},
	title = {Hyperspectral data collections at {University} of {Memphis} 2025-07-23},
	copyright = {Creative Commons Attribution 4.0 International},
	url = {https://doi.org/10.5281/zenodo.17527244},
	doi = {10.5281/ZENODO.17527244},
	urldate = {2025-11-13},
	publisher = {University of Memphis},
	author = {Watson, Thomas P},
	month = nov,
	year = {2025},
}

@misc{watson_spacecube_2025,
	address = {GitHub},
	title = {Spacecube},
	url = {https://github.com/JacobsSensorLab/spacecube-paper-release/tree/aerial-rx-paper-release},
	author = {Watson, Thomas P.},
	year = {2025},
}

@inproceedings{dastranj_remix_2025,
	address = {Charlotte, NC, USA},
	title = {{REMIX}: {Real}-{Time} {Hyperspectral} {Anomaly} {Detection} for {Small} {UAVs}},
	copyright = {https://doi.org/10.15223/policy-029},
	isbn = {979-8-3315-1328-3},
	shorttitle = {{REMIX}},
	url = {https://ieeexplore.ieee.org/document/11007863/},
	doi = {10.1109/ICUAS65942.2025.11007863},
	urldate = {2026-01-27},
	booktitle = {2025 {International} {Conference} on {Unmanned} {Aircraft} {Systems} ({ICUAS})},
	publisher = {IEEE},
	author = {Dastranj, Melika and De Smet, Timothy and Wigdahl-Perry, Courtney and Chiu, Kenneth and Bihl, Trevor and Boubin, Jayson},
	month = may,
	year = {2025},
	pages = {60--66},
}

@article{caba_concurrent_2025,
	title = {Concurrent execution of lossy compression and anomaly detection of hyperspectral images on {FPGA} devices},
	volume = {22},
	issn = {1861-8200, 1861-8219},
	url = {https://link.springer.com/10.1007/s11554-025-01692-0},
	doi = {10.1007/s11554-025-01692-0},
	language = {en},
	number = {3},
	urldate = {2026-01-27},
	journal = {Journal of Real-Time Image Processing},
	author = {Caba, Julián and Barba, Jesús and Díaz, María and Mira, José Luis and López, Sebastián and López, Juan Carlos},
	month = jun,
	year = {2025},
	pages = {112},
}

@article{melian_real-time_2021,
	title = {Real-{Time} {Hyperspectral} {Data} {Transmission} for {UAV}-{Based} {Acquisition} {Platforms}},
	volume = {13},
	issn = {2072-4292},
	url = {https://www.mdpi.com/2072-4292/13/5/850},
	doi = {10.3390/rs13050850},
	language = {en},
	number = {5},
	urldate = {2026-01-27},
	journal = {Remote Sensing},
	author = {Melián, José M. and Jiménez, Adán and Díaz, María and Morales, Alejandro and Horstrand, Pablo and Guerra, Raúl and López, Sebastián and López, José F.},
	month = feb,
	year = {2021},
	pages = {850},
}

@article{yang_onboard_2025,
	title = {Onboard {Real}-{Time} {Hyperspectral} {Image} {Processing} {System} {Design} for {Unmanned} {Aerial} {Vehicles}},
	volume = {25},
	issn = {1424-8220},
	url = {https://www.mdpi.com/1424-8220/25/15/4822},
	doi = {10.3390/s25154822},
	language = {en},
	number = {15},
	urldate = {2026-01-27},
	journal = {Sensors},
	author = {Yang, Ruifan and Huang, Min and Zhao, Wenhao and Zhang, Zixuan and Sun, Yan and Qian, Lulu and Wang, Zhanchao},
	month = aug,
	year = {2025},
	pages = {4822},
}

@article{xue_novel_2023,
	title = {A {Novel} {Method} {Based} on {GPU} for {Real}-{Time} {Anomaly} {Detection} in {Airborne} {Push}-{Broom} {Hyperspectral} {Sensors}},
	volume = {15},
	issn = {2072-4292},
	url = {https://www.mdpi.com/2072-4292/15/18/4449},
	doi = {10.3390/rs15184449},
	language = {en},
	number = {18},
	urldate = {2026-01-27},
	journal = {Remote Sensing},
	author = {Xue, Tianru and Wang, Chongru and Xie, Hui and Wang, Yueming},
	month = sep,
	year = {2023},
	pages = {4449},
}

@article{piccinini_onboard_2025,
	title = {Onboard {Hyperspectral} {Super}-{Resolution} with {Deep} {Pushbroom} {Neural} {Network}},
	volume = {17},
	issn = {2072-4292},
	url = {https://www.mdpi.com/2072-4292/17/21/3634},
	doi = {10.3390/rs17213634},
	language = {en},
	number = {21},
	urldate = {2026-01-27},
	journal = {Remote Sensing},
	author = {Piccinini, Davide and Valsesia, Diego and Magli, Enrico},
	month = nov,
	year = {2025},
	pages = {3634},
}

@article{neri_real-time_2024,
	title = {Real-{Time} {AI}-{Assisted} {Push}-{Broom} {Hyperspectral} {System} for {Precision} {Agriculture}},
	volume = {24},
	issn = {1424-8220},
	url = {https://www.mdpi.com/1424-8220/24/2/344},
	doi = {10.3390/s24020344},
	language = {en},
	number = {2},
	urldate = {2026-01-27},
	journal = {Sensors},
	author = {Neri, Igor and Caponi, Silvia and Bonacci, Francesco and Clementi, Giacomo and Cottone, Francesco and Gammaitoni, Luca and Figorilli, Simone and Ortenzi, Luciano and Aisa, Simone and Pallottino, Federico and Mattarelli, Maurizio},
	month = jan,
	year = {2024},
	pages = {344},
}

@article{chhapariya_target_2024,
	title = {Target {Detection} and {Characterization} of {Multi}-{Platform} {Remote} {Sensing} {Data}},
	volume = {16},
	issn = {2072-4292},
	url = {https://www.mdpi.com/2072-4292/16/24/4729},
	doi = {10.3390/rs16244729},
	language = {en},
	number = {24},
	urldate = {2026-01-28},
	journal = {Remote Sensing},
	author = {Chhapariya, Koushikey and Ientilucci, Emmett and Buddhiraju, Krishna Mohan and Kumar, Anil},
	month = dec,
	year = {2024},
	pages = {4729},
}

@article{su_hyperspectral_2022,
	title = {Hyperspectral {Anomaly} {Detection}: {A} survey},
	volume = {10},
	copyright = {https://ieeexplore.ieee.org/Xplorehelp/downloads/license-information/IEEE.html},
	issn = {2168-6831, 2473-2397, 2373-7468},
	shorttitle = {Hyperspectral {Anomaly} {Detection}},
	url = {https://ieeexplore.ieee.org/document/9532003/},
	doi = {10.1109/MGRS.2021.3105440},
	number = {1},
	urldate = {2026-01-28},
	journal = {IEEE Geoscience and Remote Sensing Magazine},
	author = {Su, Hongjun and Wu, Zhaoyue and Zhang, Huihui and Du, Qian},
	month = mar,
	year = {2022},
	pages = {64--90},
}

@article{scafutto_monitoring_2025,
	title = {Monitoring oil spill thickness and weathering using {UAV}-borne hyperspectral sensing},
	volume = {218},
	issn = {0025326X},
	url = {https://linkinghub.elsevier.com/retrieve/pii/S0025326X25006095},
	doi = {10.1016/j.marpolbul.2025.118134},
	language = {en},
	urldate = {2026-01-28},
	journal = {Marine Pollution Bulletin},
	author = {Scafutto, Rebecca Del'Papa Moreira and Lassalle, Guillaume and Alves, Marcos Nopper and Miranda, Lucas De Paula and Costa, Priscila Martins Oliveira Da and Souza Filho, Carlos Roberto De},
	month = sep,
	year = {2025},
	pages = {118134},
}

@article{arias_mapping_2025,
	title = {Mapping {Harmful} {Algae} {Blooms}: {The} {Potential} of {Hyperspectral} {Imaging} {Technologies}},
	volume = {17},
	issn = {2072-4292},
	shorttitle = {Mapping {Harmful} {Algae} {Blooms}},
	url = {https://www.mdpi.com/2072-4292/17/4/608},
	doi = {10.3390/rs17040608},
	language = {en},
	number = {4},
	urldate = {2026-01-28},
	journal = {Remote Sensing},
	author = {Arias, Fernando and Zambrano, Maytee and Galagarza, Edson and Broce, Kathia},
	month = feb,
	year = {2025},
	pages = {608},
}

@article{ram_systematic_2024,
	title = {A systematic review of hyperspectral imaging in precision agriculture: {Analysis} of its current state and future prospects},
	volume = {222},
	issn = {01681699},
	shorttitle = {A systematic review of hyperspectral imaging in precision agriculture},
	url = {https://linkinghub.elsevier.com/retrieve/pii/S0168169924004289},
	doi = {10.1016/j.compag.2024.109037},
	language = {en},
	urldate = {2026-01-28},
	journal = {Computers and Electronics in Agriculture},
	author = {Ram, Billy G. and Oduor, Peter and Igathinathane, C. and Howatt, Kirk and Sun, Xin},
	month = jul,
	year = {2024},
	pages = {109037},
}

@article{hasler_-it-yourself_2024,
	title = {From {Do}-{It}-{Yourself} {Design} to {Discovery}: {A} {Comprehensive} {Approach} to {Hyperspectral} {Imaging} from {Drones}},
	volume = {16},
	issn = {2072-4292},
	shorttitle = {From {Do}-{It}-{Yourself} {Design} to {Discovery}},
	url = {https://www.mdpi.com/2072-4292/16/17/3202},
	doi = {10.3390/rs16173202},
	language = {en},
	number = {17},
	urldate = {2026-01-28},
	journal = {Remote Sensing},
	author = {Hasler, Oliver and Løvås, Håvard S. and Oudijk, Adriënne E. and Bryne, Torleiv H. and Johansen, Tor Arne},
	month = aug,
	year = {2024},
	pages = {3202},
}

@article{zhao_global_2015,
	title = {Global and {Local} {Real}-{Time} {Anomaly} {Detectors} for {Hyperspectral} {Remote} {Sensing} {Imagery}},
	volume = {7},
	issn = {2072-4292},
	url = {https://www.mdpi.com/2072-4292/7/4/3966},
	doi = {10.3390/rs70403966},
	language = {en},
	number = {4},
	urldate = {2026-01-28},
	journal = {Remote Sensing},
	author = {Zhao, Chunhui and Wang, Yulei and Qi, Bin and Wang, Jia},
	month = apr,
	year = {2015},
	pages = {3966--3985},
}

@article{li_hyperspectral_2024,
	title = {Hyperspectral {Object} {Detection} {Based} on {Spatial}–{Spectral} {Fusion} and {Visual} {Mamba}},
	volume = {16},
	issn = {2072-4292},
	url = {https://www.mdpi.com/2072-4292/16/23/4482},
	doi = {10.3390/rs16234482},
	language = {en},
	number = {23},
	urldate = {2026-01-28},
	journal = {Remote Sensing},
	author = {Li, Wenjun and Yuan, Fuqiang and Zhang, Hongkun and Lv, Zhiwen and Wu, Beiqi},
	month = nov,
	year = {2024},
	pages = {4482},
}

@misc{watson_spacecube_2026,
	title = {Spacecube: {A} fast inverse hyperspectral georectification system},
	shorttitle = {Spacecube},
	url = {http://arxiv.org/abs/2601.05181},
	doi = {10.48550/arXiv.2601.05181},
	urldate = {2026-01-28},
	publisher = {arXiv},
	author = {Watson, Thomas P. and Jacobs, Eddie L.},
	month = jan,
	year = {2026},
	note = {arXiv:2601.05181 [eess]},
}

\end{document}